\documentclass[sigconf]{acmart}
\usepackage{makecell}
\usepackage{graphicx}
\usepackage{geometry}
\usepackage{subcaption}

\newcommand{\deletedcontent}[1]{}
\newcommand {\addedcontent}[1]{\textcolor{black}{#1}}

\newcommand{\StopGap}{\textsc{StopGap}~}
\newcommand{\StopGapnospace}{\textsc{StopGap}}
\newcommand{\eg}{\textit{e.g.,}~}
\newcommand{\ie}{\textit{i.e.,}~}

\newcommand{\vs}{\textit{vs}~}
\newcommand{\pquote}[1]{{\textit{\textcolor{ACMDarkBlue}{``#1''}}}}
\newcommand{\dquote}[1]{{``#1''}}

\newcommand{\FigRef}[1]{Figure \ref{#1}}

\makeatletter
\newcommand*\bigcdot{\mathpalette\bigcdot@{.5}}
\newcommand*\bigcdot@[2]{\mathbin{\vcenter{\hbox{\scalebox{#2}{$\m@th#1\bullet$}}}}}
\makeatother

\AtBeginDocument{%
  }

\makeatletter
\def\@ACM@copyright@check@cc{}
\makeatother

\copyrightyear{2025}
\acmYear{2025}
\setcopyright{cc}
\setcctype{by-nd}
\acmConference[CHI '25]{CHI Conference on Human Factors in Computing Systems}{April 26-May 1, 2025}{Yokohama, Japan}
\acmBooktitle{CHI Conference on Human Factors in Computing Systems (CHI '25), April 26-May 1, 2025, Yokohama, Japan}\acmDOI{10.1145/3706598.3714262}
\acmISBN{979-8-4007-1394-1/25/04}

\begin{document}

\title{Exploring the Design Space of Real-time LLM Knowledge Support Systems\addedcontent{: A Case Study of Jargon Explanations}}

\author{Yuhan Liu}
\authornote{Work done during an internship at Accenture Labs, San Francisco}
\orcid{0000-0001-6852-6218}
\affiliation{%
  \institution{Princeton University}
  \city{Princeton}
  \state{NJ}
  \country{USA}
}
\email{yl8744@princeton.edu}

\author{Aadit Shah}
\authornotemark[1]
\orcid{0009-0005-2655-0152}
\affiliation{%
  \institution{Ohio State University}
  \city{Columbus}
  \state{OH}
  \country{USA}
  }
\email{shah.1713@osu.edu}

\author{Jordan Ackerman}
\orcid{0009-0002-8309-4550}
\affiliation{%
 \institution{Accenture Labs}
 \city{San Francisco}
 \state{CA}
 \country{USA}
 }
\email{jordan.ackerman@accenture.com}

\author{Manaswi Saha}
\orcid{0000-0003-2981-9370}
\affiliation{%
 \institution{Accenture Labs}
 \city{San Francisco}
 \state{CA}
 \country{USA}
 }
 \email{manaswi.saha@accenture.com}

\renewcommand{\shortauthors}{Liu et al.}

\begin{abstract}
Knowledge gaps often arise during communication due to diverse backgrounds, knowledge bases, and vocabularies. With recent LLM developments, providing real-time knowledge support is increasingly viable, but is challenging due to \deletedcontent{human limitations in cognition }\addedcontent{shared and individual cognitive limitations} (\eg attention, memory, and comprehension) and the difficulty in understanding the user's context and internal knowledge. To address these challenges, we explore the key question of \textit{understanding how people want to receive real-time knowledge support}. We built \StopGapnospace---a prototype that provides real-time knowledge support for explaining jargon words in videos---to conduct a design probe study (\textit{N=}24) that explored multiple visual knowledge representation \deletedcontent{support} \addedcontent{formats}. Our study revealed individual differences in preferred representations and highlighted the importance of user agency, personalization, and mixed-initiative assistance. Based on our findings, we map out six key design dimensions for real-time LLM knowledge support systems and offer insights for future research in this space.
\end{abstract}


\begin{teaserfigure}
  \includegraphics[width=\textwidth]{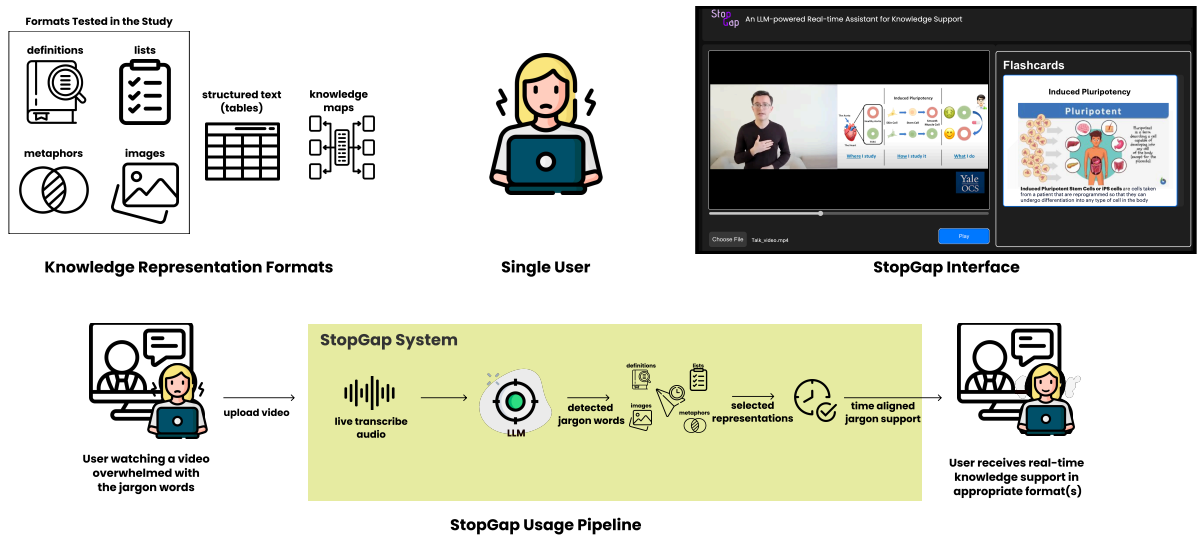}
  \caption{In this paper, we explore the design space of real-time knowledge support systems by (1) [Top-Right] building \StopGapnospace, a prototype that takes a video input and generates real-time jargon word explanations in appropriate formats. The bottom of the figure illustrates the usage pipeline of the current \StopGap prototype (2) [Top-Left] conducting a user study to test a curated set of knowledge representations from prior literature (3) mapping out a design space for such systems. This paper's focus is on single users interacting with the system.}
  \label{fig:teaser}
  \Description{Top-right is the knowledge representation formats in our design space, including definitions, lists, metaphors, images, structured text (tables), and knowledge maps. We explored the first four in this study. Top-left is the interface of the StopGap system (Figure 2). On the bottom is the usage pipeline of the StopGap system: at first, users feel overwhelmed when watching a video full of jargon words, and with the StopGap system, the video is being live transcribed, from which LLM auto-detects jargon words, select the "best" representation format and provide time-aligned jargon explanations to users. Therefore user receives real-time knowledge support in the appropriate format(s).}
\end{teaserfigure}


\maketitle

\section{Introduction}

Knowledge gaps commonly arise due to \deletedcontent{disparities }\addedcontent{differences} in individuals' educational and professional backgrounds~\cite{tichenor1970mass, w2011knowledge, gaziano2017knowledge, donohue1973mass}. These gaps are sometimes blatant (\eg a totally unfamiliar word) but can also remain undetected because individuals tend to overestimate the extent of their own knowledge~\cite{kruger1999unskilled, sloman2018knowledge}. In the context of cross-domain communication---where participants possess expertise in different fields---such knowledge gaps can be even more pronounced. This is in part due to the tendency of individuals to assume that others share their knowledge, leading them to employ domain-specific jargon that may not be comprehensible to those outside their field~\cite{nickerson1987people, nickerson1999we, misleads2005projective}. \addedcontent{Prior research has pointed out the knowledge asymmetry in the use of jargon as a key challenge in knowledge-based communication, even impeding the daily operation of organizations~\cite{patoko2014impact, friedman2011you, eppler2007knowledge}}.

Traditional knowledge support systems have predominantly provided asynchronous assistance (\ie not real-time) as there were previously core limitations in the speed and quality of information retrieval ~\cite{ginsburg1999annotate, said1987kess, liu2005task}. Although recent advancements in \addedcontent{information retrieval technologies have made real-time support feasible, these solutions remain largely restricted to ad-hoc databases~\cite{dumais2004implicit, zhao2018calendar, livne2014citesight, milde2016ambient}. While Large Language Models (LLMs) offer the potential for real-time knowledge generation that goes beyond the specific knowledge warehouse,}\deletedcontent{Large Language Models (LLMs) have enabled the possibility of real-time knowledge generation,} limitations\addedcontent{and variation} in human cognitive capacity still pose significant difficulties for delivering effective real-time knowledge support~\cite{wada2000basic, marois2005capacity}. 

Knowledge representations play a vital role in building knowledge support systems ~\cite{o1998enterprise, ginsburg1999annotate} and can be designed in human or machine-readable formats. Knowledge representation formats can include structured text, tables, images, stories, and visual metaphors~\cite{eppler2007visual}. However, there is an absence of formative research examining users' perceptions and preferences regarding these formats, particularly how they impact cognitive load in real-time settings. Cognitive load can be thought as the mental effort to process information during a task~\cite{hart1988development}. In this paper, we aim to investigate: \textbf{\textit{how do we design knowledge representations for a real-time setting that does not overwhelm the user?}}

To investigate this question, we conducted a qualitative design probe study (\textit{N=}24) to elicit feedback and opinions on the appropriate knowledge representations for a real-time context. Based on the prior work about visual knowledge representations\addedcontent{~\cite{eppler2007visual}} and our testing of LLM's generation capability, we employ four formats in explaining jargon words in this paper: definition, text-based metaphor, list, and image. For the design probe, we built \StopGapnospace, a real-time knowledge support system that is an audio \& video-playing tool integrated with an AI assistant that explains technical words or jargon using different representation formats. The qualitative study was a mixed design, having (1) a within-subjects experiment, where we\deletedcontent{study} \addedcontent{compared} the \StopGap experience of participants' \addedcontent{with their} current knowledge gap-filling method, and (2) a between-subjects factor where participants were divided into audio-only and audiovisual groups to assess the impact of potential distractions or benefits from \StopGap \vs video visuals. \addedcontent{To ensure the consistency in the study material seen by participants, we generated the representation in advance and displayed it to participants synchronously while they used the system}. Quantitative data were collected through quizzes on the content of the presentations viewed by participants. We also collect NASA Task Load Index (TLX) surveys to assess participant cognitive load~\cite{hart1988development, pachunka2019natural}. We analyzed the interview transcripts using thematic analysis and identified key themes.

Our findings show that participants perceive the real-time knowledge support provided by \StopGap as useful and appreciate it because of the automation, \addedcontent{just-}in-time support, and the multiple knowledge representation formats. The data we collected during the study shows that \StopGap enhanced people's understanding of the\deletedcontent{presentation} \addedcontent{video content} while not observably increasing their cognitive load. Though there is no universal answer to \dquote{\textit{what is the best knowledge representation format?}} as good representations can be contextual and personal, participants shared their perceptions and insights on different formats. Lastly, participants explicitly and implicitly indicated in the study that they value user agency, personalization, and mixed-initiative assistance in real-time knowledge support systems. Based on our analysis, we propose a design space for building real-time knowledge support systems.  

In summary, our contributions include:
\begin{itemize}
    \item Qualitative findings from a design probe study on people's perceptions of enhancing real-time knowledge support with the consideration of the user's cognitive load
    \item A design space for building real-time knowledge support systems
    \item Insights into the challenges and opportunities of leveraging LLMs to generate real-time knowledge assistance, as well as an in-depth discussion of the trade-offs between user agency and automation in knowledge support systems.
\end{itemize}

\section{Related Work}

\subsection{Knowledge \& Knowledge Gaps}
Knowledge can be measured as both dichotomous and continuous measures~\cite{gaziano2017knowledge}. The former can be described as \dquote{\textit{knowledge of}} something, which is the awareness of something while the latter one can be described as \dquote{\textit{knowledge about}} something, which is the in-depth or mechanistic information about a topic~\cite{park1940news, donohue1973mass}. Moreover, knowledge gaps emerge when differences in information acquisition occur between groups, often driven by factors such as education, socioeconomic status, and other disparities~\cite{tichenor1970mass, w2011knowledge}. These gaps are typically more pronounced as complex information is learned across various topics and domains by different individuals~\cite{gaziano2017knowledge, w2011knowledge}. Prior studies have pointed out that knowledge gaps represent a key challenge in communication and can cause relational tension between domain experts and decision makers~\cite{mengis2007integrating, szulanski2000process}.

Sometimes we are explicitly aware of knowledge gaps, and sometimes we are not. In cross-domain communication, ignorance of knowledge gaps may be more frequent. On one hand, people are more likely to assume others have knowledge if they possess it themselves. In other words, they may use a lot of jargon in domains they are more familiar with during communication because they tend to assume other people also process this knowledge~\cite{tullis2023curse, nickerson1999we, nickerson1987people, misleads2005projective}. On the other hand, self-awareness about one's own knowledge can be elusive. Studies have shown that humans often have an illusion that they have \dquote{knowledge about} things when in reality they may only have surface-level \dquote{knowledge of} the target phenomena (\eg Do you know how a toilet works? - Yes; Can you explain it? umm... No)~\cite{sloman2018knowledge}. In other words, we are not always reliable reporters of our own knowledge and if we incorrectly think we know something (\eg how a toilet works or what a word means), we may miss opportunities to address that knowledge gap, especially when only relying on user-initiated remedies (\eg Googling, asking an AI). 

\addedcontent{Among communication problems caused by knowledge gaps, over usage of jargon words has been identified as a primary factor that hinders effective communication, especially in modern organizations where employees often come from diverse cultural and educational backgrounds~\cite{patoko2014impact, friedman2011you, eppler2007knowledge}. In this paper, we aim to explore methods for closing knowledge gaps in real-time settings (\eg cross-domain conversations),} \deletedcontent{focusing on explaining jargon words}\addedcontent{using jargon explanations as the first step of exploration. }

\subsection{Knowledge Management and Support Systems}
\subsubsection{Components in Knowledge Support Systems}
Knowledge support systems are designed to facilitate the acquisition of specialized information mostly within organizations and are composed of five key components~\cite{ginsburg1999annotate, o1998enterprise}:
\begin{enumerate}
    \item \textbf{\textit{Knowledge warehouse}}: to store knowledge.
    \item \textbf{\textit{Knowledge search and discovery mechanisms}}: to tease knowledge out of data warehouses.
    \item \textbf{\textit{Knowledge representation}}: to create formats of the knowledge that are human-readable or machine-readable.
    \item \textbf{\textit{Knowledge filtering tools}}: to establish minimum credibility of the knowledge.
    \item \textbf{\textit{Knowledge visualization models}}: to present the knowledge to the end users via visualization (\eg webpages).
\end{enumerate}

Earlier knowledge support systems, like \addedcontent{Knowledge Engineering Support System (KESS)~\cite{said1987kess}}, Annotate~\cite{ginsburg1999annotate}, and K-support~\cite{liu2005task}, ranging from expert database building to organizational productive pinching, face similar challenges concerning how to effectively store and retrieve data~\cite{ginsburg1999annotate}. The speed of information retrieval previously constrained the scenarios where knowledge support could be applied, necessitating asynchronous systems. \deletedcontent{However, with recent advancements in distributed databases and large language models (LLMs), large-scale data storage and high-speed information retrieval and generation are no longer practical limitations. In addition, the contextual understanding LLMs afford enables newly flexible and fast queries, making real-time knowledge support increasingly feasible. }\addedcontent{Though information retrieval speed has been accelerated in recent years, making real-time support more practical, the application is restricted to specific scenarios and narrowly scoped databases. These systems primarily focus on task-specific environments rather than real-time environments (\eg cross-domain conversations). Prior systems have focused on utilizing state-of-the-art retrieval algorithms to enhance user productivity by augmentation with relevant information such as calendar information and message history for email management~\cite{dumais2004implicit, zhao2018calendar}, pulling academic reference recommendations during manuscript writing~\cite{livne2014citesight}, and analyzing speech streams and querying relevant documents in live interaction~\cite{milde2016ambient}. Despite their ability to deliver real-time knowledge support, these systems rely heavily on pre-configured databases. Users must explicitly define the types of knowledge they require in advance. However, as discussed earlier, in real-time communication there are many scenarios where knowledge gaps emerge unpredictably, making it impractical to predefine or preconfigure the necessary knowledge warehouses. Recent HCI studies have shown that LLMs can generate in-situ information in real-time in video conferencing applications~\cite{liu2023visual}, accessibility systems~\cite{valencia2023less, liu2024human} and mixed-reality~\cite{de2024llmr}. In this study, we aim to explore a real-time knowledge support solution based on LLMs, which delivers flexible and fast responses without the need to define a knowledge repository in advance.}

\subsubsection{Knowledge Representation Formats}
As O’Leary~\cite{o1998enterprise} points out, knowledge representations can be either human-readable or machine-readable. But how do we create human-readable knowledge representations? Prior research by Eppler and Burkhard~\cite{eppler2007visual} proposes a framework for building visual representations. The framework addresses critical factors; what type of knowledge is visualized? Why should that knowledge be visualized? For whom is the knowledge visualized? In which context should the knowledge be visualized? And how can the knowledge be represented? The authors also provide an overview of different visualization formats, from which we take inspiration for our study. Those formats include:
\begin{enumerate}
    \item \textbf{\textit{Structured text and tables}}: textual items with different colors, fonts, and font sizes integrated into visual structures like tables and trees. 
    \item \textbf{\textit{Mental images and stories}}: visuals or narratives that conveys ideas and enhance understanding
    \item \textbf{\textit{Heuristic sketches}}: drawings that highlight problem-solving potentials or capture people's mental models. 
    \item \textbf{\textit{Conceptual diagrams}}: schematic depictions to structure information and illustrate relationships.
    \item \textbf{\textit{Visual metaphors}}: graphic or symbolic representation that bridges something familiar to something new, which can also improve memorability ~\cite{worren2002theories}.
    \item \textbf{\textit{Knowledge maps}}: graphic formats that follow cartographic conventions to reference relevant knowledge~\cite{eppler2004making, burkhard2005knowledge}.
    \item \textbf{\textit{Interactive visualizations and animations}}: interactive, computer-supported visualizations that enable users to engage with and manipulate information, promoting both knowledge transfer and creation.
\end{enumerate}

\deletedcontent{However, these were not made with the real-time context in mind.} In our paper, we focus on the core question of exploring the design of knowledge representations that work effectively in a real-time context. \addedcontent{We specifically study the formats \#1, \#2 and \#5. It is important to note that the knowledge representation formats proposed by Eppler and Burkhard are not restricted to jargon explanations but to all types of knowledge support. We use jargon explanations as the first step towards investigating the space.} 


\subsection{Real-time Support in \addedcontent{Computer-Mediated} Communication}
\deletedcontent{Over the past decades, researchers have explored how to provide communication support in real-time settings. This support includes facilitating communication in both physical and knowledge-based tasks. While the type of task may vary, the goal remains largely the same: offering assistance to help individuals complete their tasks while minimizing the distraction caused by the support system~\cite{kosch2017one, scholz2013concept, aseniero2020meetcues, das2022cannot, song2021online}. To tackle the challenge of distractions in task-support systems, researchers have implemented several strategies, such as removing distraction sources, reducing the harmful effects caused by distraction, etc.~\cite{son2023okay, aseniero2020meetcues, das2022cannot, song2021online, marlow2016taking, avrahami2016supporting}.}

\addedcontent{Prior HCI literature has investigated how to facilitate real-time communication either via speaker/presenter support (\eg physical gesture-based tools ~\cite{saquib2019interactive}), listener support (\eg LLM-based conversation support tools~\cite{liu2023visual}), and group support (awareness tools~\cite{dimicco2004designing, dimicco2007impact}). However, the ability to digest information is a key challenge in real-time communication, especially where the support could be distracting if not designed well.}

\addedcontent{Marois and Ivanoff note that the major bottleneck of information processing lies in human's ability to perceive, hold in mind, and act upon the visual information received~\cite{marois2005capacity}. For ease in processing information, managing distraction is a key consideration for real-time systems, unlike asynchronous knowledge support systems.  Previous real-time support systems focused on helping individuals complete their tasks via distraction minimization techniques (\eg adjust screen lighting ~\cite{aseniero2020meetcues}) or by adapting to their cognitive capability~\cite{wada2000basic}}. While distraction may have benefits depending on the cause and the extent~\cite{benbunan2012measurement, twyman2020too}, for real-time systems it may be more of a hinderance. Therefore, in this paper, we focus on studying how to design real-time knowledge support that does not overwhelm the user in the moment.




\section{StopGap System}
\begin{figure*}[t]
  \includegraphics[width=\textwidth]{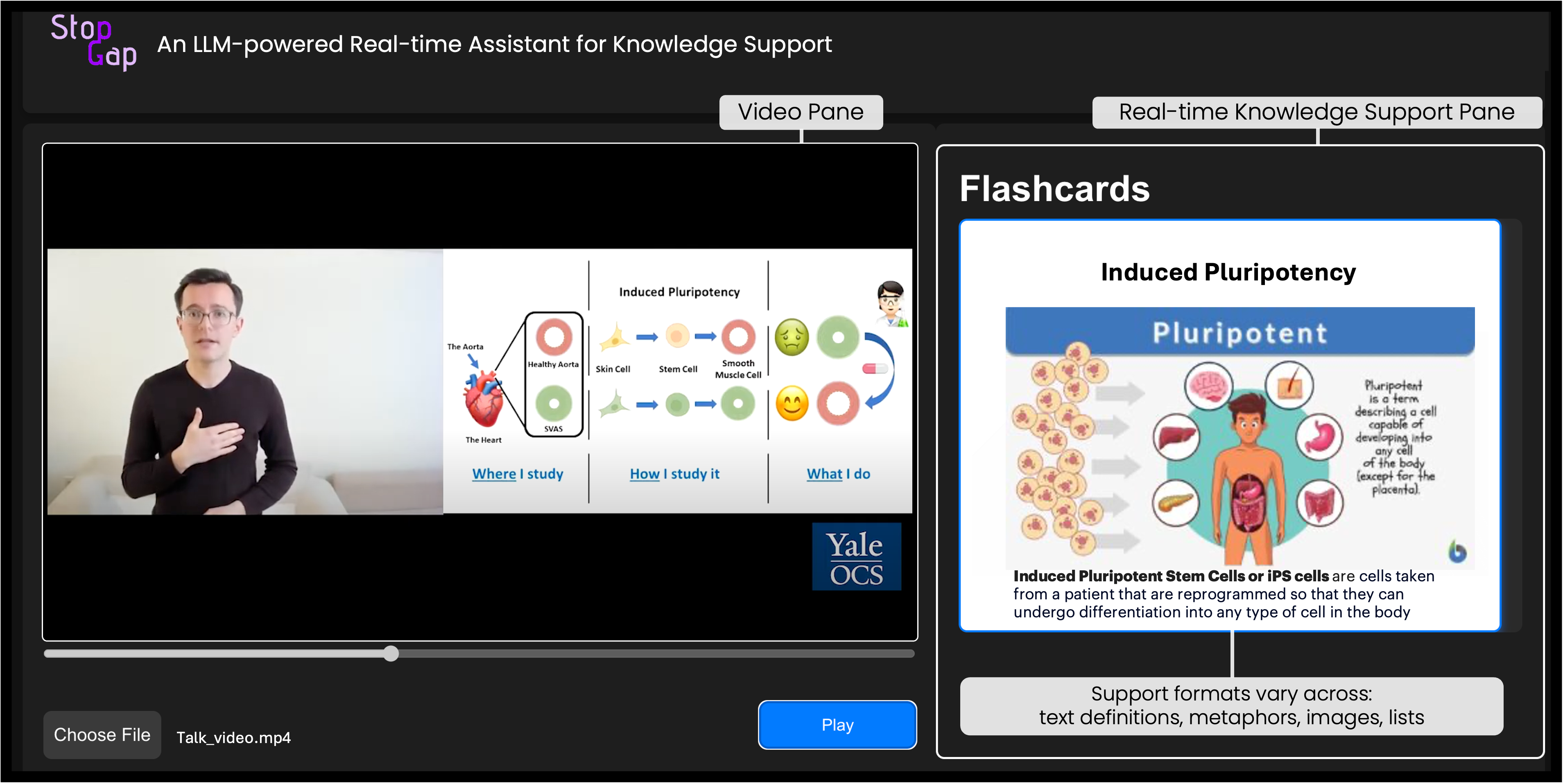}
  \caption{StopGap Prototype. To use this system, a user uploads a video. The system auto-transcribes and batch-identifies jargon words. The system then displays the real-time knowledge support in a side panel as flashcards once the user starts playing the video. The flashcards scroll up automatically for every new jargon word.}
  \Description{StopGap Interface: Video is playing on the left side. The system will automatically detect jargon words from the presenter's speech and generate corresponding explanations in the flashcards}
  \label{fig:stopgapsys}
  \Description{The interface of the StopGap system. On the left is a screenshot of a video when the speaker is introducing induced pluripotency. On the right is a flashcard explaining induced pluripotency with images and textual definitions.}
\end{figure*}

\StopGap (\FigRef{fig:stopgapsys}) is a prototype designed to provide real-time knowledge support by explaining jargon words from technical videos. Our work builds on previous research in augmented communication, which explored the potential for augmentation and its implementation~\cite{liu2023visual} and visual knowledge representations~\cite{eppler2007visual}, which illustrated the commonly seen knowledge representation formats and examples. 

\subsection{System Description}
The \StopGap system takes in a video as an input and generates and displays knowledge support in different formats in real time. To reduce the participant's time to learn the system, we designed the \StopGap interface to resemble a video player with only four components.

\begin{enumerate}
    \item \textit{\textbf{Video Panel}}. The uploaded video is displayed here. When the user uploads a video, the system auto-transcribes and generates the list of time-stamped jargon words. Once the batch is created, the video controls are enabled.
    \item \textit{\textbf{Real-time Knowledge Support Pane}}. \textit{Flashcards} are used to contain the real-time support in a chosen format, shown in the side panel. For this study, we chose to show only one format at a time. However, future versions could explore combining formats for better understanding as shown in \FigRef{fig:stopgapsys}. The flashcards auto-scroll with every new jargon word. The user cannot see the previously shown flashcards.
    \item \textit{\textbf{Video Controls}}. The user can play and pause at any time. However, they cannot go back/forward.
    \item \textit{\textbf{Progress Bar}}. Similar to any video player, it shows how far the video has been played.
\end{enumerate}

\textbf{Note}. For the study, the LLM-based knowledge support generation was disabled to ensure every participant saw the same support for each jargon word.

\noindent

\subsection{Design Considerations}

\subsubsection*{Build Flexibility} We chose to build the system from scratch rather than creating plugins for existing communication tools, due to their low flexibility in creating and displaying knowledge representations. 

\subsubsection*{Knowledge Representation Formats} 
\begin{figure}[t]
\includegraphics[width=\linewidth]{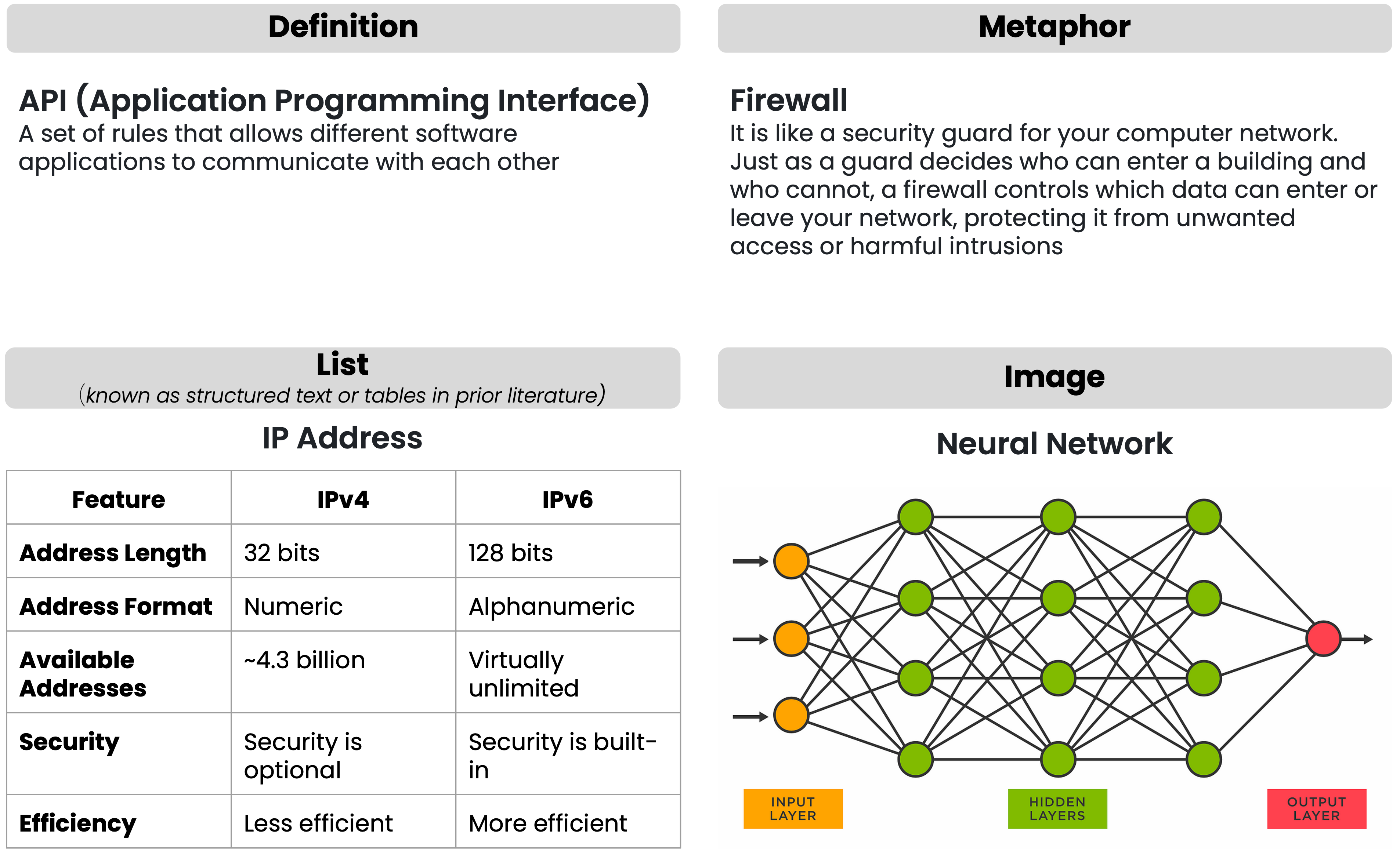}
\caption{Chosen Knowledge Representation Formats.}
\label{fig:knw_format}
\Description{Examples of the four knowledge representation formats we explored in this study. a) Using definition to explain API (Application Programming Interface): a set of rules that allows different software applications to communicate with each other. b) Using metaphor to explain firewall: It is like a security guard for your computer network. Just as a guard decides who can enter a building and who cannot, a firewall controls which data can enter or leave your network, protecting it from unwanted access or harmful intrusions. c) Using list to explain IP Address (comparing IPv4 and IPv6): IPv4 is 32-bit long, has numeric format and IPv6 is 128-bit long, and has alphanumeric format. There are around 4.3 billion available IPv4 addresses while IPv6 addresses are virtually unlimited. The security of IPv4 is optional while the security of IPv6 is built-in. IPv6 is more efficient than IPv4. d) using image to explain the neural network, from left to right are: input layer, hidden layers, and output layer. Each layer has multiple nodes and nodes in nearby layers are inter-connected.}
\end{figure}

We selected the knowledge formats based on prior work by Eppler and Burkhard ~\cite{eppler2007visual}, who proposed seven formats: structured text and table; mental images and stories; heuristic sketches; conceptual diagrams; visual metaphors; knowledge maps; interactive visualizations and animations. To better support our focus of real-time jargon word explanations, we added jargon \dquote{definitions} as an additional text-based format. Based on initial testing, we found that LLM-generated knowledge representation in formats like heuristic sketch, conceptual diagram, knowledge map, and image was often confusing. However, acknowledging the importance of image-based formats and the necessity to explore people's perception of such formats for real-time contexts, we opted to source Internet images from Google Search. Animations were excluded due to concerns about comprehending complex information in a short time. Consequently, we selected four formats for the study (\FigRef{fig:knw_format}): \textit{definitions} $\cdot$ \textit{text-based metaphors} $\cdot$ \textit{images} $\cdot$ \textit{lists}. Lists are renamed from structured text/tables to avoid confusion for participants.

\subsubsection*{Knowledge Support Placement} We designed the user interface with flashcards on the right side to resemble mainstream communication software such as Google Meet, Teams, Zoom, and YouTube, where features like chat, participants list, and related content recommendations typically appear. 

\subsubsection*{Real-time Support Frequency} To accommodate cases where jargon words appear too closely together (\eg in the same sentence), we hard-code a 5-second minimum interval between flashcards, a decision informed by internal testing within our research team. \addedcontent{If there are two jargon words in the same sentence, the first card will remain displayed on the screen for 5 seconds until the second one is presented.}

\subsubsection*{Automation and User Agency} The system is almost fully automated because we want to closely mimic real-time communication. Thus, disabling users from moving back and forth while watching a video. In addition, this design choice was motivated by the need to assess whether users could grasp jargon meanings without needing to revisit the content on the flashcards. However, we make an exception for participants to pause. The ability to pause allows users to practically extend the default 5-second flashcard interval to give them more time to process.

\subsection{System Implementation}
\StopGap (\FigRef{fig:stopgapsys}) is an LLM-based web application designed to facilitate the understanding of unfamiliar video content by explaining jargon words in real time. The application takes in MP4 video files as input and the system generates knowledge support in one of four representation formats. It utilizes React for the front-end interface and Node.js and Express.js for the back-end services, along with Whisper~\cite{radford2023robust} for speech-to-text transcription and GPT-4 for jargon detection and explanation \cite{achiam2023gpt}. Upon uploading a video, the system first uses Whisper to transcribe the audio into text. This text is then analyzed by GPT-4 to identify and extract key jargon terms. Each run of the system may produce slightly different outputs. To ensure the consistency of knowledge support elements across participants in this study, a single representative set of generated jargon words (for each video) was selected, along with their associated timestamps. As the video plays, identified jargon terms are displayed in real-time at their corresponding timestamps, allowing users to grasp specialized concepts as they appear. Each time a new video is uploaded, the system resets, transcribing the new video, detecting the jargon, and displaying the latest jargon flashcards in sync with the video.

\section{Design Probe Study}

We conducted a think-aloud qualitative study to investigate our RQ on understanding what knowledge representations work in a real-time setting that does not distract or overwhelm the user. We used \StopGap as the design probe. 

\subsection{Study Methodology}
\begin{figure}[t]
\includegraphics[width=\linewidth]{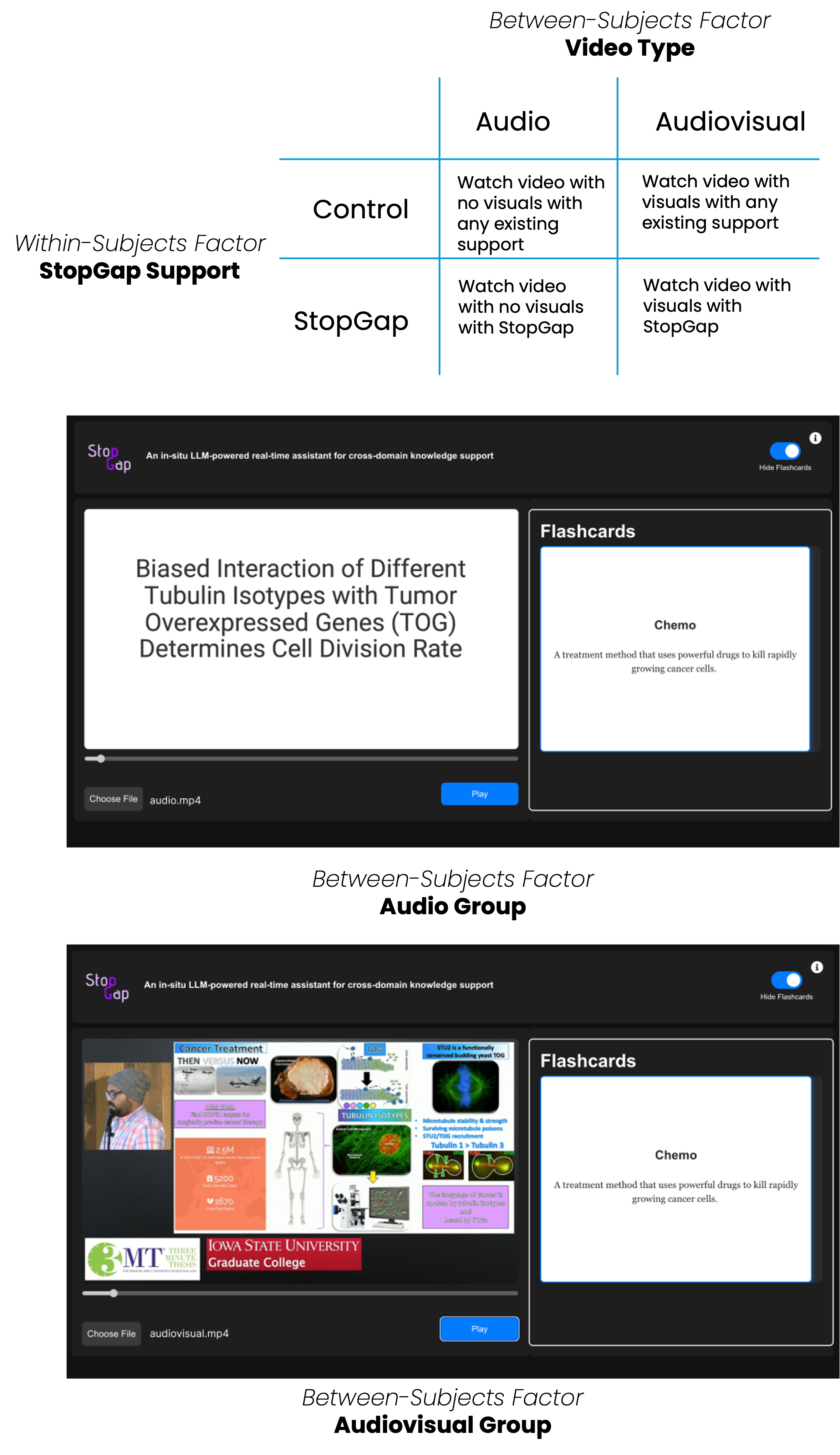}
\caption{Mixed Study Design}
\label{fig:study_design}
\Description{Figure to illustrate the study design, for the within-subjects factor, we have control and Stopgap task, for between-subject factor, we have audio and audiovisual group. The difference between audio and audio visual group is we masked the visuals in the audio group and only show participants the title of the presentation in interface.}
\end{figure}

We took a mixed-design study approach, with a within-subjects factor (\StopGap support) and a between-subjects factor (video type) (\FigRef{fig:study_design}). Through three pilot studies, we iteratively settled on the study parameters: the basic study design (mixed), the number, and the type of representation formats to be shown per jargon word. See Appendix ~\ref{appendix:pilot_study} for more details on the pilot study iterations.

We divided the participant pool into two groups using the between-subjects factor. The first group watched a video with visuals kept intact from the source video (\textit{audiovisual} group) and the second group heard the talk without the visuals from the source video (\textit{audio} group). This was done to isolate the potential interaction of visual distraction from the \StopGap knowledge support with the video visuals. Across the two groups, each participant did two tasks: (a) Control task, where they watched a video without \StopGap assistance, but were free to use any existing support tools such as Google, ChatGPT to look up terms (b) \StopGap task, where participants watched a different video in the \StopGap tool without access to any other support tools. We counterbalanced the task order across participants. Each task was followed by a short quiz testing participants' understanding of the video content and a survey measuring their self-reported cognitive load using the NASA TLX questionnaire~\cite{hart1988development}. 


To ensure consistency of knowledge support experience across participants, we curated a set of jargon words to show support for and created explanations using all four representation formats for each jargon word. During the \StopGap task, participants were presented with support in a single knowledge representation format, selected by the researcher as the most relevant for the given word context. However, we conducted a post-task reflection exercise, where the researcher walked through the entire list of jargon words and the associated explanations in the four formats. Participants were asked to choose the most useful format and explain why. 

Finally, we concluded the study with a debrief interview asking participants to compare their user experience in the two tasks and elicit their perceptions and opinions on the system, including likes, dislikes, and design suggestions. Study sessions were \addedcontent{around} 60 minutes \addedcontent{(\textit{Mean=}59.03 minutes, \textit{SD=}4.95 minutes)} and were conducted on Zoom except one was in person. We audio and screen captured the participant sessions. Participants \addedcontent{were recruited through internal communication channel} and were compensated with \$75/hour. The study protocol is attached in Appendix \ref{appendix:study_protocol}. \addedcontent{The study is approved by the internal ethics and legal board}.

\subsection{Study Materials}

\subsubsection{Study Video Selection} 
To ensure the videos across the two study tasks were at the same level of difficulty, we conducted an evaluation on a set of 3-minute videos from CHI conference talks as well as \dquote{Three Minute Thesis} repository. We conducted two rounds of video evaluations by people unfamiliar with the video topic. Evaluators were asked to watch the selected videos, review jargon words detected by an LLM, and evaluate the difficulty of understanding the video content. Based on the evaluations, we selected two award-winning videos from the \dquote{Three Minute Thesis} repository (Table \ref{tab:videos}). Both videos are related to Cell Biology and are from the same speaker. On average, evaluators rated the difficulty level of understanding of the two videos 3.2 and 3.4 respectively (5-point Likert Scale, with 5 being very difficult). To further quantitatively compare the difficulty, we transcribed the video and calculated the average sentence length, clause density, and Flesch-Kincaid readability score ~\cite{flesch2007flesch} (Table \ref{tab:videos}).

\begin{table*}[t]
    \centering
    \caption{Characteristics of Study Task Videos}
    \label{tab:videos}
    \begin{tabular}{p{4.2cm}p{5cm}p{5.4cm}}
    \hline
        ~ & \textbf{ \href{https://vimeo.com/showcase/8974954/video/640414115}{Video 1}} & \textbf{ \href{https://vimeo.com/showcase/9978088/video/770798360}{Video 2}} \\ \hline
        \textbf{Title} & Remote Spatial Coordination within Cytoskeleton Regulates Timing of Cell Division & Biased Interaction of Different Tubulin Isotypes with Tumor Overexpressed Genes (TOG) Determines Cell Division Rate \\ \hline 
        \textbf{Video Length} & 177s & 170s \\ 
        \textbf{\#LLM-detected Jargon words} & 9 & 9 \\ 
        \textbf{Average Sentence Length} & 17.14 & 18.79 \\ 
        \textbf{Clause density} & 1.67 & 2.29 \\ 
        \textbf{Flesch-Kincaid Readability} & 11.5 & 10 \\ \hline
    \end{tabular}
\end{table*}

\subsubsection{Jargon Word Selection and Explanation Generation} We designed \StopGap to generate real-time knowledge support using an LLM. However, the LLM output cannot be guaranteed to be the same even using the same prompt~\cite{OuYang2023LLMIL}. To control the consistency in the materials presented to participants, we hand-curated and defined the jargon words and explanations in the study. We ran the system 10 times and created a list of jargon words that appeared more than 5 times, each video ended up with 9 jargon words. \addedcontent{Note that there is no overlap between the sets of jargon words identified in the two videos}. We then created explanations for each jargon word across the four formats. For text-based explanations \addedcontent {(definition, metaphor, and list)}, we used ChatGPT 4o. For image-based explanations, we searched the jargon word on Google Images and found an appropriate image. The appropriateness was evaluated by a domain expert, who also further refined the explanations during the pilot study.

\subsection{Participants}
We recruited 24 people (\textit{M=}11, \textit{F=}13) aged 18--54. On a 5-point Likert scale measuring familiarity with biology (5 being Very Familiar), all participants were in 1-3 range (\textit{Mean}=1.875, \textit{Median}=2, \textit{SD}=0.68). Most participants were early career corporate professionals ranging from 0-5 years of experience, except for three with 10+ years. They had diverse educational backgrounds, including Business Administration, Human \& organizational development, English Literature, Statistics, Data Science and more. Participants were recruited through a call posted within the organization's communication platform. \addedcontent{The organization requires cross-domain communication on a daily basis making it a good candidate for recruiting participants}. More participant details are in Appendix ~\ref{appendix:participants}.



\subsection{Analysis}
We conducted thematic analysis~\cite{braun2006using} on 24\deletedcontent{hour-long} \addedcontent{60-minute long} audio transcripts, focusing on participants' perceptions of different knowledge representation formats and their opinions on \StopGap like real-time knowledge support systems. We first categorized high-level code groups according to our study protocol, and then developed open codes based on four transcripts. Two experts in the research team validated the codebook by having several peer-debriefing discussion sessions. The codes were updated based on the discussions. The primary coder then coded the rest of the transcripts with the second researcher as the reviewer. Finally, we grouped similar codes and extracted themes according to our research questions~\cite{mcdonald2019reliability}. The codebook is available in the Supplementary Materials.

\section{Findings}
In this section, we report our findings on the real-time knowledge support provided by \StopGap from three perspectives: (1) perceived usefulness and cognitive load, (2) participants' preferences and perceptions of the various representation formats utilized in the system, and (3) participants' desired modes of interaction with the system. 


\subsection{\addedcontent{Perceived Usefulness and Cognitive Load}}

\subsubsection{Though their reasons varied, participants reported\deletedcontent{perceiving} the real-time knowledge provided by \StopGap as useful}
We asked participants to rate the usefulness of the knowledge support they received, on a scale of 1-5, 5 being very useful; we observed an average of 3.67 across 24 participants (\textit{SD=}0.76, \textit{Median=}4). \deletedcontent{Participants noted several considerations including immediacy, automation, and the multiple information formats they saw in the system.} \addedcontent{In addition to the perceived usefulness, } we assessed participants' understanding of the video content using both quizzes and self-reported \deletedcontent{questions}\addedcontent{scores} (in-study questionnaire attached in the Appendix \ref{appendix:questionnaire}) as shown in Table \ref{tab:understanding_results}. \deletedcontent{
For quizzes, participants performed slightly better in the \StopGap task compared to the control task, with an average score of 3.42 (\textit{Median=}4, \textit{SD=}0.83) \vs 3.25 (\textit{Median=}3, \textit{SD=}0.79) out of four questions. Additionally, on a scale of 1-5, we asked participants to rate their familiarity with the jargon used before and after both tasks, where 1 indicated no familiarity and 5 indicated very high familiarity. The average difference in familiarity after completing the \StopGap task was 1.04 (\textit{Median=}1, \textit{SD=}0.69), while the corresponding value for the control task was 0.88 (\textit{Median=}1, \textit{SD=}0.85). Participants were also asked to rate how successfully they believed they understood the jargon, with higher scores indicating greater perceived understanding. On average, participants reported higher confidence in their understanding during the \StopGap task (\textit{Mean=}13.13, \textit{Median=}13.5, \textit{SD=}3.77) compared to the control task (\textit{Mean=}12.71, \textit{Median=}13, \textit{SD=}4.77). Overall, these } \addedcontent{Our} results suggest that participants performed slightly better and felt more confident in their understanding during the task with real-time jargon support compared to the control task. \addedcontent{Though the t-tests showed no statistical significance across the three metrics, participants elaborated on how StopGap improved their understanding of the content compared to their current methods}.

\deletedcontent{We define the \dquote{usefulness} of the knowledge support provided by \StopGap by comparing it with what participants currently use to fill the knowledge gap. Therefore} \addedcontent{During the control task,} we asked participants about their current methods for filling knowledge gaps. They mention simultaneously watching the video and searching through search engines (\eg Google, Bing), and querying LLMs (\eg GPT, Llama). Other noted methods were hybrid: taking notes while watching the video and highlighting keywords to figure out the meanings afterward. When asked to compare the user experience of using their current method and \StopGapnospace, participants reported \deletedcontent{several benefits of the current methods they use including,} \addedcontent{the main benefit of their current methods as the} tailoring to their needs. \deletedcontent{because}\addedcontent{For example,} when people prompt ChatGPT, they can give specific instructions on what kind of information they can provide and what kind of information they want to receive, which is not enabled in \StopGapnospace: \pquote{Some of the language (\StopGap used) was not as easy as I would have liked it to be. It should be like if I was in high school or at a similar level. If I asked ChatGPT that, for a definition, it would maybe give it a little bit more simple} (P3). 

\addedcontent{In the design of \StopGapnospace, we prioritized automation instead of customization to ensure the knowledge support can be in real-time}\deletedcontent{In fact, in the design of \StopGapnospace, we trade such customization into the automation of the jargon word detection and explanation format selection to achieve high immediacy}, which some participants appreciated. For example, P13 considered that manually querying LLMs was time-consuming while in \StopGapnospace, the knowledge support is triggered automatically and is integrated together with video playing: \pquote{What I like about \StopGapnospace, is that you don't have to type like: `Please explain to me ...' That is where I switch my focus, and lose a couple of seconds. It automatically scans the text for certain terms and tries to explain.} (P13). Furthermore, P17 compared the effort needed for getting variety in Google search and in \StopGapnospace: \pquote{I think being able to have a variety of different formats come into your view, that's super helpful without having to go to a variety of different sources on Google} (P17). Participants also liked the automatic knowledge gap detection because they sometimes lack the context to initiate the search even though they want to, as P14 mentioned: \pquote{If there was specific terminology that I wasn't familiar with, I didn't know how to spell it correctly, I wasn't able to look it up. It was hard for me to bridge that information gap and understand what exactly this is.} (P14). 

\begin{table*}[!ht]
\caption{Summary of Quiz Scores and Self-Reported Understanding Across Tasks}
\label{tab:understanding_results}
    \centering
    \begin{tabular}{lllllll}
    \hline
        ~ & \multicolumn{2}{c}{Quiz scores \addedcontent{[0-4]}} & \multicolumn{2}{c}{\makecell{Self-reported difference of \\familiarity after \\and before task \addedcontent{[-4-4]}}} & \multicolumn{2}{c}{\makecell{Self-reported score on \\ successfulness of \\understanding \addedcontent{[1-21]}}}  \\ \hline
        Task & Control & \StopGap & Control & \StopGap & Control & \StopGap\\ \hline
        Mean & 3.25 & 3.42 & 0.88 & 1.04 & 12.71 & 13.13 \\ 
        Median & 3 & 4 & 1 & 1 & 13 & 13.5 \\ 
        SD & 0.79 & 0.83 & 0.85 & 0.69 & 4.77 & 3.77 \\ \hline 
        \multicolumn{7}{c}{\makecell{Note: for all metrics, a higher score means a better understanding}}
    \end{tabular}
\end{table*}

\subsubsection{\addedcontent{\StopGap provides real-time knowledge support without causing overwhelming cognitive overload}\deletedcontent{Quantitative Results for Participants' Understanding and Cognitive Overload}}

\deletedcontent{
For quizzes, participants performed slightly better in the \StopGap task compared to the control task, with an average score of 3.42 (\textit{Median=}4, \textit{SD=}0.83) \vs 3.25 (\textit{Median=}3, \textit{SD=}0.79) out of four questions. Additionally, on a scale of 1-5, we asked participants to rate their familiarity with the jargon used before and after both tasks, where 1 indicated no familiarity and 5 indicated very high familiarity. The average difference in familiarity after completing the \StopGap task was 1.04 (\textit{Median=}1, \textit{SD=}0.69), while the corresponding value for the control task was 0.88 (\textit{Median=}1, \textit{SD=}0.85). Participants were also asked to rate how successfully they believed they understood the jargon, with higher scores indicating greater perceived understanding. On average, participants reported higher confidence in their understanding during the \StopGap task (\textit{Mean=}13.13, \textit{Median=}13.5, \textit{SD=}3.77) compared to the control task (\textit{Mean=}12.71, \textit{Median=}13, \textit{SD=}4.77). Overall, these results suggest that participants performed slightly better and felt more confident in their understanding during the task with real-time jargon support compared to the control task.}

\begin{figure}[hbt!]
\centering
\includegraphics[width=\linewidth]{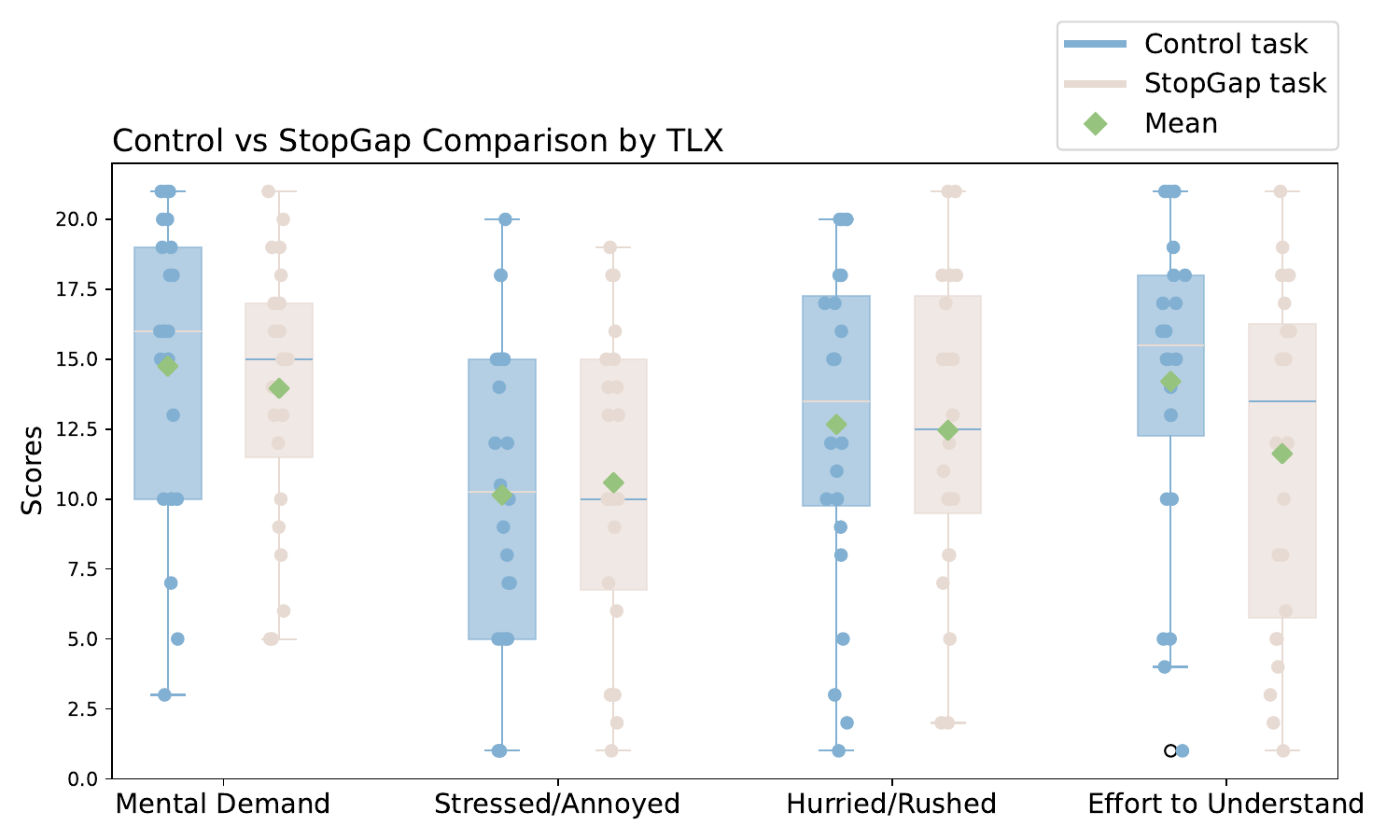}
\caption{Comparison of participants' task load index of control task and \StopGap task}
\label{fig:tlx}
\Description{Comparison of participant's task load index of control task and StopGap task. In terms of mental demand and stressed or annoyed, StopGap task shows more concentrated distribution. For hurried or rushed and effort to understand, control task seems to be more concentrated in distribution. The average and median are very close among two tasks.}
\end{figure}

We also explored the cognitive load participants experienced using the NASA Task Load Index (TLX)~\cite{hart1988development} (Question details in the Appendix ~\ref{appendix:questionnaire}). \addedcontent{As shown in Table \ref{tab:tlx_results} and Figure \ref{fig:tlx}, }\deletedcontent{
Results are shown in Figure \ref{fig:tlx} and Figure \ref{fig:tlx_comparison}.}there is no clear trend indicating that the real-time support increased user cognitive load.
To break down different aspects of the cognitive load, the mean and median of \deletedcontent{each index} \addedcontent{mental demand, feeling stressed or annoyed and feeling hurried or rushed} for the \StopGap task and the control task are very similar, indicating our system didn't overwhelm our participants.\deletedcontent{Participants did report that \StopGap made them feel stressed, annoyed, hurried, and rushed. However, the system may have also lowered their cognitive demand and effort to comprehend the video (jargon in particular). Participants in the audiovisual group reported feeling less hurried and rushed and less mentally strained than in the audio group. We noticed in both tasks that audiovisual group participants felt more stressed and annoyed, which indicates there is more distraction when multiple information sources are presented to people simultaneously. However, in \StopGapnospace, the effort required to understand the video in \StopGap is less for the audiovisual group compared to the audio group while in the control group, the result shows the opposite.} \addedcontent{Participants even reported lower scores in the effort to comprehend the video content when using \StopGap and attributed it to the help provided by \StopGap counteracted the distraction caused by getting confused by jargon words. As P6 stated:\pquote{It helps me track (the video content) and not get distracted by the big words that I wasn't familiar with.} P6 reported much lower scores in mental demand (19 for the control task, and 6 for \StopGap task), stressed or annoyed (11 for the control task and 7 for \StopGap task), and hurried or rushed (12 for the control task and 2 for \StopGap task). In the comparison between audio and audiovisual groups (see Figure \ref{fig:tlx_comparison} for details), we observed in general a lower score in the audiovisual group than in the audio group. This counterintuitive finding indicates that more visual information does not necessarily mean more distraction. Participants illustrated that the information provided by \StopGap doesn't bother them as they can selectively digest the needed information: \pquote{I don't think it's (having visuals) distracting. It's giving me just enough, and I can choose as a consumer of what information I want and there's not too much stuff where I feel bogged down.} (P5 who was in the audio group but mentioned wanting more than just audio). P2 echoed this and pointed out he would turn to \StopGap for help when only needed: \pquote{If you come in with zero information. It's overwhelming because it keeps prompting new information. But in a work setting, I wouldn't come in with zero knowledge. It would be more like an aid for understanding just those few things that I don't understand.}} (P2)

\addedcontent{After performing a t-test, we found no significant difference in participants' perceived cognitive load when using \StopGap. As shown in previous work, confusion caused by unfamiliar knowledge can lead to distraction~\cite{dharmasena2021nexus}.  In our study, some participants reported that \StopGap’s real-time support helped mitigate this distraction, preventing them from feeling overwhelmed by the additional information. This suggests that real-time knowledge support systems have the potential to assist users in managing cognitive load, though the extent of their effectiveness may vary across individuals.}

\begin{table*}[!ht]
\caption{Summary of Task Load Index of Different Groups Across Tasks}
\label{tab:tlx_results}
    \centering
    \begin{tabular}{lllllllll}
    \hline
        ~ & \multicolumn{2}{c}{Mental Demand}  & \multicolumn{2}{c}{Stressed or Annoyed}  & \multicolumn{2}{c}{Hurried or Rushed} & \multicolumn{2}{c}{Effort to Understand} \\ \hline
        Task & Control & \StopGap & Control & \StopGap  & Control  & \StopGap & Control & \StopGap \\ \hline
        All Mean & 14.75 & 13.96 & 10.15 & 10.58 & 12.67 & 12.46 & 14.21 & 11.63 \\ 
        Median & 16.00 & 15.00 & 10.25 & 10.00 & 13.50 & 12.50 & 15.50 & 13.50 \\ 
        SD & 5.20 & 4.66 & 5.68 & 5.44 & 5.90 & 5.46 & 5.71 & 6.13 \\ \hline
        Audio Mean & 14.92 & 14.25 & 10.88 & 10.50 & 12.83 & 11.92 & 14.33 & 12.33 \\ 
        Median & 16.00 & 15.50 & 12.00 & 10.00 & 13.50 & 10.50 & 16.50 & 13.50 \\ 
        SD & 5.92 & 3.84 & 6.22 & 3.90 & 5.84 & 5.20 & 6.91 & 5.69 \\ \hline
        Audiovisual Mean & 14.58 & 13.67 & 9.42 & 10.67 & 12.50 & 13.00 & 14.08 & 10.92 \\
        Median & 15.00 & 15.00 & 8.50 & 12.00 & 12.50 & 15.00 & 15.00 & 12.50 \\ 
        SD & 4.64 & 5.52 & 5.26 & 6.83 & 6.20 & 5.89 & 4.52 & 6.71 \\ \hline
        \multicolumn{9}{c}{\footnotesize Note: for all metrics in the table, the higher the index is, the higher the cognitive load the participant experiences}
    \end{tabular}
\end{table*}

\deletedcontent{To sum up, our quantitative results indicate that \StopGap may enhance understanding of the content while seeming to not increase cognitive load. From the interviews, we learned that the integrated experience and automated knowledge gap detection offset the distractions caused by introducing a new information medium. As prior work has highlighted, it is inevitable to have distractions sometimes in building communication-facilitating systems, but the negative impact can be mitigated through the benefit it brings~\cite{son2023okay}. }

\subsection{\addedcontent{Preferences for Representation Formats}}

We find that no one format can be sufficient in explaining a jargon word or a concept in all cases. The optimal knowledge representations will likely vary depending on factors such as the nature of the content, the concept being communicated, and the individual (\eg cognition, learning habits, and cultural background).  

\subsubsection*{Definitions}
Definitions are good for getting straightforward ideas and reinforcing existing knowledge in a short amount of time. However, it can get too technical, containing extra jargon words, which hinders understanding as P24 noted: \pquote{The definition was less useful for me because of my background knowledge. I'm not as aware of what tubulin or eukaryotic cells are. So I would probably need additional double clicks of definitions, to know what that means.}---explaining the difficulty in understanding a definition with more jargon words, in this case, using ``tubulin'' and ``eukaryotic cells'' to explain ``Microtubules''.

\subsubsection*{Metaphors} 
Metaphors are good for complicated and uncommon words as they usually use simple language. Participants indicated they believed metaphors would enhance their long-term memory of the jargon word as they connect the jargon word with objects and activities people are familiar with. For example, P16 finds the easy wording of the metaphor very useful: \pquote{I'll choose metaphor because the explanation is very clear for a layman to understand what exactly it is} (P16). However, generating metaphors is hard as it is challenging to find precise analogies. Whether the analogy is accurate, or whether it can resonate with its audience depends on what they’re more familiar with, which is decided by their daily practice and backgrounds. A bad analogy can be misleading, and even make understanding harder. In our study, P13 stated the misleading caused by the analogy he didn't understand: \pquote{I am not sure if it is the right metaphor, because, like malfunctioning factory machines that churn out harmful products at an uncontrollable rate when we are talking about factory machines. The first thing that comes to my mind is like: ‘Okay, we are getting a lot of waste. We will think about utilizing that waste later. Let's just throw the waste in the window. Maybe after some time, the machines will start outputting something good.’ So it does not underline the importance.}---P13's understanding of the jargon from the metaphor is wrong because the malfunctioning factory metaphor misled him. In addition, the generated metaphor is sometimes not concise enough to support immediate information. It can also be used to supplement more straightforward formats such as definitions and images. 

\subsubsection*{List} The perceptions around this format were more polarized. Some think the lists contain too much information which makes them feel overwhelmed while others think more information provides a good extension to better comprehend the content. P19 noted the difficulty in comprehending too many words in lists: \pquote{The list just seems like a lot of words. It's like a lot of different keywords just thrown at you.} (P19). In contrast, P3 mentioned \pquote{(The list) It's comprehensive, and I think it does a good job of explaining what Chemo is like, and what the uses, benefits, and limitations are. If I just wanted to learn as much as I could about Chemo, then that's what I would want to look at}. This not only varies based on personal perceptions but also applies to different words. For example, for words people have a basic understanding of or words that can be inferred from root words, participants are leaning toward using a list format to gain more information, while for uncommon words, the list format causes too much information overload. Additionally, participants reported the bullet points in the list are good for catching key points and are consistent with their common reading habits and existing mental models: \pquote{Since we work in consulting, we're very familiar with reading like bullet point structured things so that just like automatically follows some kind of system in my head} (P24). 

\subsubsection*{Image} 
Image format provides distinct affordances compared to text-only representations, and it has been shown in prior research to be a good learning medium~\cite{bobek2016creating}. Aligned with previous research, many participants noted that they identify as visual learners and found images useful for quickly digesting information and also leaving a deeper impression. Like P4 has told us: \pquote{I think it's easier to understand and remember. Oftentimes, when you're recalling information, you're not really recalling words in your brain, but you can recall an image.} (P4). However, most participants found the image format to be too vague. Specifically, the image could have been applied to explain other things than the jargon word, so participants missed the point. For example, p7 thought the image to explain \dquote{Chemo} could also be used to explain \dquote{hospital}. Finally, the semantic relevance of the image to the video content was an issue, more pronounced in the image format compared to the other formats. For example, P16 was confused about whether the image to explain the word `Fluorescently' was actually the figure used in the speaker's research: \pquote{When you project an image, we really are not sure whether it is referring to the study or an example that he (the speaker) is showing. So I was not able to get that.} (P16). We attribute the above-mentioned problems to the source of the image. The images shown to participants in the study were from Google search, which may be graphs and figures from other research papers or scientific reports. 

\textit{Summary}. There are pros and cons for each format. Choosing when to use a specific format to explain jargon depends on the context as well as on people's preference and their familiarity with the jargon word. In some cases, different formats can be complementary to each other. In our study, participants suggested combining different formats as the best explanation for some jargon words, especially for lists and images. They found the missed information in those two formats could be completed by definitions or metaphors: \pquote{A combination of the definition, and the list would be helpful where, like having the definition, and then maybe calling out like the benefits and the limitations specifically.} (P14). 

\subsection{\addedcontent{Preferred Interaction Models}}
\subsubsection{Participants seek greater agency in real-time knowledge support systems}
In our study, participants explicitly or implicitly indicated that they wanted more agency in the \StopGap system. The agency is specifically about how they want to interact with the system in two ways: the choice of the format and the control of the flow of flashcards.  

Participants expressed wanting more agency in deciding the representation formats in the system, especially picking up the format that fits their knowledge base: \pquote{Maybe all four sections that we saw could be on the right. And you could just select, like the metaphor or the definition, or the picture so it wouldn't be like overload, but you would know that you could click for more. I think that could be beneficial.} (P6). Another level of agency they want is when the system-provided knowledge support fails to make them understand, they have control to get alternatives. For example, P19 suggested having a switching arrow: \pquote{If there was a way to switch between the formats. That would also be helpful. I'm envisioning a slideshow where you can just use the left arrow, right arrow, or something.} (P19) Participants also wanted to control the time they saw a specific flashcard. This is particularly needed for participants to connect different jargon words. As a participant described when they would want to revisit a flashcard: \pquote{I see myself getting confused like: `Is this similar to the Tubulins that were just mentioned? Are they different types of Tubulins? Maybe more callbacks to other terms in the flashcard'.}---P24 was talking about microtubules, which are polymers of the tubulin.  

\subsubsection{Participants value personalization in tailoring knowledge assistance}

Gaining user agency in the system cannot be achieved without personalization, as people have different preferences in terms of the level of agency. 

We learned that the jargon word detection without considering personal needs can add extra distraction to people as they may expect a flashcard to show up but they don't because the system did not recognize it as a jargon word: \pquote{When he was like mentioning specific terms, I was expecting to a flash card to pop up. So I was almost kind of like watching it and waiting for that to happen. And then not fully focusing on the video} (P24). Similarly, if a flashcard carries a word they already know, it is adds distraction as people's attention may be automatically attracted by what pops up on the screen: \pquote{At our level, it's pretty obvious what therapeutic would mean in this situation. So I felt like diverting my gaze to it was like `Oh, I understand this. I shouldn't focus my attention on this. I should just keep listening to the lecture'. (...) Looking at too simple of a word distracts from the depth of the video.} (P7). Furthermore, in terms of what specific personalization they want to have in jargon word detection, P13 suggested referring to the English language assessments where education websites test people's vocabulary to decide which level they are at: \pquote{It would be great if \StopGap had some system that would estimate the level of expertise of the audience. And by using that methodology to determine which words need to be explained.} (P13). 

Besides detecting jargon words, participants mentioned deciding explanation format should also be based on personal preferences and learning habits: \pquote{Each person learns differently. So probably you should have all 4 available. But let the user customize what works best for them. Since we all learn differently.} (P10). However, none of the participants gave specific details on what kind of personalization and what level of personalization they want to have.

\subsubsection{Participants wish to have mixed-initiative assistance in real-time knowledge support}

When asked about how they deal with knowledge gaps in their daily communication, participants' current methods are all user-initiated, which requires the knowledge gap to be identified first, and then filled. In developing \StopGapnospace, we intentionally made the system fully automated. As reported before, some participants liked the automation, which saved them time and effort. The system has no doubt, to some extent, added another source of information when the user is actively engaged in real-time communication: \pquote{It was also a little distracting that I became more focused on looking at the \StopGap than paying attention to the video. So I had to remind myself to keep focusing on the video itself and not just reading the \StopGapnospace.} (P21). To mitigate the distraction by multiple information sources, participants suggested adding a trigger mechanism of the knowledge support so that they can toggle it off when they feel too overwhelmed by the information overload: \pquote{I think having more manipulatable design would be helpful. Instead of just having it on the side. It's like having the slides on your own side when a presenter is talking like being able to flip through them on your own.} (P24). Participants also suggested integrating the knowledge support in subtitles or closed captions which will be triggered through hovering or clicking on certain words: \pquote{Imagine the systems interface that there are subtitles, and even without stopping subtitles, I can hover the mouse cursor over the long, unknown words, such as photosynthesis. I move my mouse over that word, the word gets highlighted, I click on it, and it puts a flashcard on a stack so that I don't have just one flashcard on my right side. I have a stack of flashcards.} (P13). 

\textit{Summary}. The balance between self-initiation and automation is hard to reach in system design because individual users have different preferences in terms of the freedom they wish to have in controlling the system. Personal preferences may also vary due to different times and scenarios. We will further discuss mixed-initiative assistance in the next section.
\section{Discussion}
Drawing from prior literature on design spaces of real-time augmented communication~\cite{liu2023visual} and our study findings, we first map the design space of real-time knowledge support systems. We then summarize the design implications of generating appropriate real-time knowledge representations and highlight insights on how to personalize knowledge support based on people's preferences and knowledge bases. Finally, we discuss the dilemma of granting user agency in interaction with the system and keeping the convenience and immediacy of fully automated systems.


\begin{figure*}[hbt!]
\centering
\includegraphics[width=\linewidth]{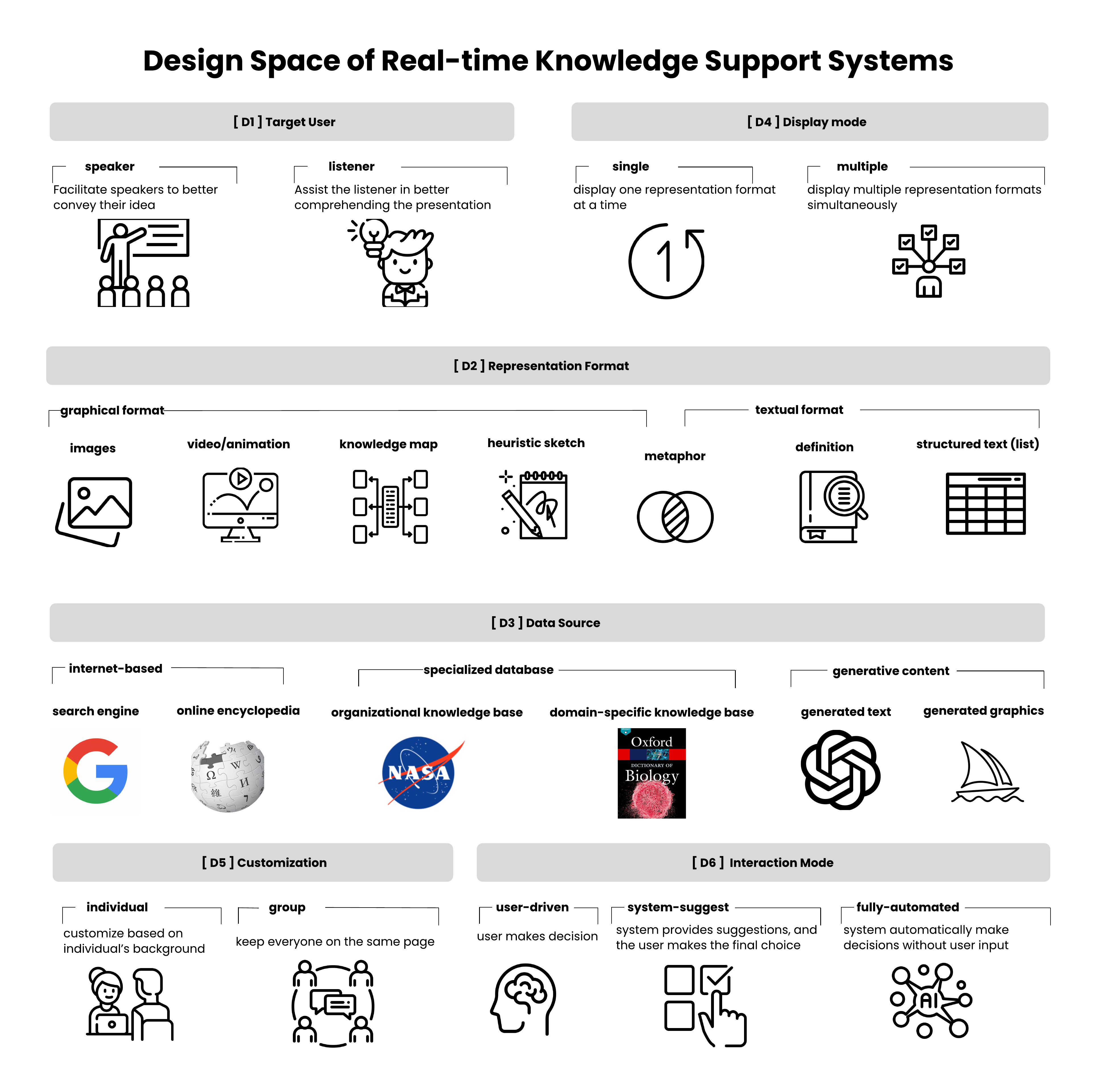}
\caption{Design Space of Real-time Knowledge Support Systems}
\label{fig:design_space}
\Description{Design space of building real-time knowledge support system. We proposed 6 dimensions. The figure is the same as section 6.1 but with more icons of the dimensions.}
\end{figure*}

\subsection{Design Space for Real-time Knowledge Support Systems}
Using design space analysis methods~\cite{card1990design, card1991morphological} and building on our qualitative data and prior research, we identified six key design dimensions, detailed in Figure \ref{fig:design_space}.

\subsubsection{D1: Target User}
Real-time knowledge support systems can be used in two fundamental scenarios, for speakers and for listeners. For speakers, this kind of support can help them prepare content (\eg speech, presentations) and ensure the content is audience-appropriate -- this is a form of automated perspective-taking where presenters can explore which words or concepts may be considered jargon to different audiences. \addedcontent{In other words, speakers can rehearse their presentation with the system and revise their presentation content based on different target audiences. Additionally, our work extends beyond video-watching contexts, aiming to generalize real-time support to broader communication scenarios. In these settings, speakers can utilize knowledge representations as supplementary materials to enhance the clarity and effectiveness of their presentation, as explored in prior research~\cite{hu2023thingshare, liu2023visual}. While the primary focus of this paper is on designing knowledge representations—which might initially seem centered on the listeners' perspective—we argue that the quality of a speaker's presentation significantly impacts the listener's experience. Including speakers as target users is therefore essential for designing effective knowledge support systems.} For listeners, this kind of support can provide assistance for listeners to quickly better comprehend the video content. 

\subsubsection{D2: Representation Format}
Derived from classical literature ~\cite{eppler2007visual} and our findings, we identify two categories of knowledge representation formats as being the most appropriate: \textit{\deletedcontent{image-based}\addedcontent{graphical format}}, which includes images, \addedcontent{heuristic sketches, graphical metaphors\footnote{Note that metaphors can be either graphical or textual}, knowledge maps} and animations, and \textit{\deletedcontent{text-based} \addedcontent{textual format}}, which includes definitions, list (structured text/tables), \addedcontent{and symbolic metaphors}. Both categories convey different elements of knowledge support that are easy to consume in a real-time context.

\subsubsection{D3: Data Source}
Traditional knowledge management systems focus on providing the organizational level of knowledge support for internal use~\cite{o1998enterprise, ginsburg1999annotate, said1987kess}. However, our research expands to a wider variety of knowledge bases. Data sources are not limited to organizational data but can range from Internet-based sources such as search engines (\eg{Google and Bing}) and online encyclopedias (\eg Wiki-pedia)
to specialized databases that have higher credibility than others when the application is limited to specific organizations or domains. The third data source is AI-generated content. Recent research has shown that generative AI can be used to comprehend the data and transform it into contextualized knowledge~\cite{beheshti2023empowering}. We found that LLM-generated text-based knowledge support was perceived as useful. However, participants want the knowledge support to be closely connected to the communication context, which cannot be provided by the other two data source types. AI-generated text, images, and videos can help bridge this gap by augmenting, customizing, or contextualizing content,  although specific practical support needs to be further proven. 

\subsubsection{D4: Display Mode}
This dimension refers to the number of formats to be shown to the user. In our study, we gained insights into participants' perceptions of the utility of different formats for jargon explanations. Each format demonstrated unique strengths and weaknesses. Combining different formats for knowledge support can offer good design opportunities as long as the cognitive load to digest information can be managed effectively. 

\subsubsection{D5: Customization}
Our findings show that participants highly value customization of the real-time knowledge support. Tailoring knowledge support to an individual's existing knowledge base (\eg educational and cultural background) as well as capabilities (\eg learning habits and styles, reading level) can significantly enhance its effectiveness. Furthermore, for group settings, customization would need to consider all the individuals and their existing knowledge bases and mental models and find the right balance across the diversity. Further research is needed for group knowledge support systems.

\subsubsection{D6: Interaction Mode}
Our findings indicate that participants desire user agency across various dimensions. We identify three potential interaction modes between the user and the system. The first is a fully \dquote{user-driven} mode, where users initiate knowledge support or choose their preferred format options. Second is a \dquote{fully-automated} mode, as implemented in the \StopGap prototype, where the system autonomously provides knowledge support without user input. And the third is a human-intervention mode \dquote{system-suggest}, where the system offers suggestions, but the user makes the final decision.

\subsection{Tradeoff between LLM-generated Knowledge Support and Information Credibility} 

Participants expressed the need for in-context information in a real-time setting. For example, for the system to connect current jargon word explanation to something that was said earlier in that session. This requires rapid analysis and memory needs, both of which current LLMs are capable of handling to generate contextually relevant knowledge representations.
%
%
%
%
However, the credibility of the output remains a concern due to inconsistencies and the inherent limitations in the reliability of LLM-generated content~\cite{andrus2022enhanced, ye2023assessing, wang2023survey}. During the study, participants also raised concerns about the legitimacy of the jargon word explanations they saw: where does the explanation come from, and how do we ensure the explanation is accurate? Given the issue of LLM hallucinations remains unresolved~\cite{zhang2023siren}, we propose the following mitigation strategies with the consideration of balancing the need for contextually relevant knowledge support\addedcontent{, for not only jargon word explanation generation, but also for other type of knowledge representation generation}: 
\begin{enumerate}
    \item \textit{Enhancing the transparency of data sources}. For example, adding citations to the generated knowledge, which has been proven capable with Retrieval-Augmented Language Models or Reward Models in training ~\cite{guu2020retrieval, menick2022teaching, huang2024training}. This can increase the knowledge support's trustworthiness and fulfill some users' desire for deeper, extended knowledge by allowing them to explore the underlying data sources. 
    \item \textit{Adding layers for contextual verification and tiering the information delivery}. Recent research has investigated how to distinguish whether LLMs are generating false information and how people perceive LLMs expressing uncertainty of their outputs ~\cite{kim2024m, azaria2023internal}. Therefore we advocate for adding a verification layer in knowledge support systems and delivering the information based on relevance to the communication context. The verification would use technologies to cross-validate the truthfulness of the generated knowledge representation and output confidence score. Users can select the confidence score threshold based on their trust in the LLM-generated content. 
\end{enumerate}

\subsection{Trade-off between User Agency and Automation}
In our study, participants expressed their desire for user agency in the system, particularly in jargon detection, as they are most aware of their own vocabulary and knowledge bases. While we value user control in system design, prior research also shows that the self-perceptions on topical knowledge may not always be accurate reflections~\cite{cole2010self}. For example, in our study, one participant confused \dquote{pathogenesis} with \dquote{photosynthesis} and only recognized it when prompted by flashcards. This example highlighted the necessity of automated knowledge gap detection\addedcontent{, not limited to jargon words, but anytime where the user lacks certain knowledge}. The challenge of automated knowledge gap detection lies in how the system gains a thorough picture of people's personal knowledge base without the user indicating everything they know or do not know~\cite{guo2023personalized}. The opportunity we identified is to build algorithms that test people's knowledge through a small set of questions and infer how familiar they are with certain topic areas, concepts, or terms. However, unlike user profiles in recommendation systems, knowledge bases vary significantly even among individuals with similar educational and professional backgrounds~\cite{Rescher2020}. \addedcontent{Recent work by Guo et al explored supervised and prompt-based solutions for personalized jargon detection, offering insights in integrating personal knowledge into jargon detection, but it's still within CS domain ~\cite{guo2023personalized}.} While this paper emphasizes the need for user agency and personalized support in real-time knowledge systems, its main focus is addressing knowledge gaps without overwhelming users. The role of user agency and personalization requires further research to explore how AI can enhance human cognition by balancing these elements with automated support.





\subsection{Generalizing the Design Space from Jargon Explanations to Other Types of Knowledge Gap Filling}
From our findings and prior work, we identified two key layers in designing real-time knowledge support systems: knowledge gap detection and knowledge gap filling. In our proposed design space, D1, D5, and D6 are directly related to the detection and D2, D3, and D4 are related to the knowledge gap filling. The insights we discussed in Sections 6.2 and 6.3 are not only limited to jargon explanations. We note that jargon is just one type among different knowledge categories. Previous work in knowledge engineering has categorized knowledge into four types: \textit{object knowledge} (knowledge about \dquote{things} and concepts of the domain), \textit{performance knowledge} (description of abilities or potential behaviors of an object), \textit{event knowledge} (recognition of a certain combination of object knowledge and performance knowledge occurs) and \textit{meta knowledge} (knowledge about knowledge such as knowing where to locate given knowledge)~\cite{andersen1996knowledge}. This study focused primarily on object knowledge, offering design implications to mitigate knowledge asymmetry arising from differing knowledge bases. Given that prior research highlights the interconnectedness of these knowledge types, we argue that \StopGap can be generalized to broader knowledge support contexts. Specifically, we envision future systems that incorporate an additional layer capable of breaking down performance knowledge and event knowledge into object knowledge to facilitate general knowledge support. For example, the ability of an aircraft to fly can be decomposed into object knowledge about wings, engines, and aerodynamic surfaces, which \StopGap can effectively present. We highlight the opportunity for future research to explore real-time knowledge breakdowns by leveraging diverse data sources. 

\addedcontent{In conclusion, in this paper, we present the core tenets of designing real-time knowledge support systems that can be used across knowledge types. We recognize that the cognitive load required to process information may vary across these types (and individuals) and would need further investigation for identifying relevant design dimensions to support them. We leave this extended design space exploration for future work.}

\section{Limitations and Future work}
\addedcontent{First, though the knowledge representation formats proposed by Eppler and Burkhard are not limited to explaining jargon words~\cite{eppler2007visual}, the users' cognitive load is directly related to jargon words. The cognitive load in digesting other types of knowledge (\eg performance knowledge about how to do something) might be different. Future research should address the quantification of cognitive load in relation to various knowledge gap-filling activities.} Our quantitative results suggest the potential for building real-time knowledge support systems without imposing additional cognitive load on users. While the limited sample size prohibits drawing definitive conclusions, our qualitative insights guide us on the design of such knowledge representations. Future work could focus on a large-scale study to empirically evaluate the impact of real-time knowledge support on a user's cognitive load. Additionally, we focused on exploring people's perceptions of the different knowledge representation formats; hence, using pre-defined knowledge representations to ensure consistency across participants. We refrained from directly using LLM-generated outputs or automatically retrieving Google Search images to ensure the generated representations were accurate and appropriate. Exploring the potential of LLMs in generating real-time knowledge representations presents an intriguing avenue for future work, as it may enable more flexibility and scalability in the design of such systems. We leave evaluating the reliability of such LLM-generated knowledge representations for future work.  
\section{Conclusion}
In this paper, we explored the design of knowledge representations that do not overwhelm the user in a real-time knowledge context. We conducted a design probe study (\textit{N=}24) with the \StopGap system, which integrates real-time transcription and LLM-driven jargon detection to deliver knowledge support for understanding specialized video content. Participants validated the usefulness of the real-time support provided by \StopGap and shared their insights on knowledge representation formats and how they prefer to interact with such real-time assistance. Study findings suggest future real-time LLM knowledge support systems should:
$\bigcdot$ tailor knowledge representations to user background (\eg profession, expertise)
$\bigcdot$ implement strategies to enhance the credibility of LLM-generated knowledge support
$\bigcdot$ balance user agency with automation.

\section{Disclosure of the usage of LLM}
We used ChatGPT (GPT4o model\cite{achiam2023gpt}) to facilitate the writing of this manuscript. The usage includes:
\begin{itemize}
    \item Turn Excel format tables into latex format tables
    \item Correct grammar mistakes and spelling
    \item Polish the existing writing by prompts like "Find me a synonym of X", "What is the noun/adjective form of X" and "Shorten this sentence without changing its content".
\end{itemize}


\bibliographystyle{ACM-Reference-Format}
\bibliography{0_reference}

\appendix

\appendix
\section{Pilot Study}
\label{appendix:pilot_study}
We conducted three pilot studies with different participants: one with a stakeholder who doesn't have any HCI background, one with a domain expert in the HCI field, and one with a domain expert working in the intersection with HCI and biomedical engineering. We refined our study design based on the results and feedback from these three pilot studies. 

The first pilot study compared user experience watching video without the \StopGap system, with the \StopGap system showing one format for each jargon word and with the \StopGap system showing all four formats for each jargon word. The participant indicated that the system is overwhelming even seeing only one knowledge representation format, let alone with all knowledge formats. Thus, we refined our study design to compare user experiences with and without the \StopGap system and have a reflection task later to investigate people's perceptions and opinions of different knowledge representation formats. We used the revised study protocol for the second pilot study. 

In the second pilot, the participant noticed a significant added cognitive load caused by the visuals in the video shown in addition to the \StopGap support. To better isolate the sources of distraction from the video \vs from the knowledge support, we decided to add a between-subject factor where half of the participants would be asked to view a pre-processed video with all visuals being removed from the source video. In other words, they only hear audio from the presentation without any visuals except the title. We used this study design to conduct a third pilot study with an expert in biomedical engineering. 

The third pilot focused on polishing the materials, especially the jargon word explanations across the four formats. The main goal of the user study is to understand people's perceptions and preferences in receiving real-time knowledge support, not the ability of LLM to provide such support. We consulted a domain expert on refining the jargon word explanation instead of leaving the explanation as it was originally generated. All explanations used in the study are shown in Appendix ~\ref{appendix:explanations}.


\section{Study Protocol}
\label{appendix:study_protocol}
\textbf{Semi-structured Interview Session}

\textbf{[5 min] Introduction}

Thanks for coming today. Before we start the session, let me tell you about our project. This project intends to build an AI assistant to facilitate cross-disciplinary communication in a real-time context. For example, when you are listening to project presentations from people whose backgrounds are different from yours, you may find several terminologies confusing, which hinders you from fully understanding the context. To solve this issue, we designed and developed the StopGap prototype, which can detect and explain jargon words automatically in real-time settings. 

The goal of this study is to explore the best way to provide real-time support and better understand your preferences. The study will be in three parts:
\begin{enumerate}
    \item The first part will be a task where we’ll show you an Audiovisual and you can use your preferred method to figure out the meaning of jargon words and Audiovisual content. It will be followed by a quiz testing your understanding and a survey measuring your cognitive load.
    \item The second part is watching another Audiovisual with the help of StopGap. You’ll be asked to finish a similar quiz and survey afterward.
    \item Finally, the last part is an interview study, where questions will be to understand your perception and opinions about StopGap. For example, we are interested in knowing – your likes, dislikes, and design ideas to improve it. Also, knowing your perceptions about the utility of the tool. 
\end{enumerate}

The whole study session should take about 60-70 minutes. Your data will be kept anonymous. We will be audio/Audiovisual recording. For the Audiovisual recording, you can turn off your camera if you don’t want your face being captured. You have the right to stop participating in the study at any time. Before we begin the interview, we need you to sign the consent form.
Are there any questions?

[researcher starts recording]

I have just started the recorder and we will begin the interview. Please feel free to say whatever is on your mind and ask me questions at any time. Are you ready to begin?

\textbf{[10-15 min] Control Task: Watch Audiovisual Without StopGap}

\textit{Goal: Demonstrate experience without StopGap system}

In this part, your task is to understand the content of a three-minute Audiovisual using any method you prefer. Please note that you have up to 8 minutes, including the Audiovisual length, to pause and look up any additional information to enhance your understanding, though this is optional. Afterward, there will be a short quiz, followed by a survey measuring your cognitive load. 

[3-8 min] [researcher gives remote control to the participant to play the Audiovisual, participant watches with preferred method] [researcher sets up a timer]

Q. On a scale of 1 to 5, with 1 being very easy and 5 being very difficult, how would you rate the difficulty of understanding this Audiovisual? 

Q. Could you please describe the method you use to help you figure out the content of the Audiovisual? How did the method you use [adjust based on the method participants use] impact your understanding?

I just sent you the link to the quiz and the survey, you have up to 7 minutes to finish the first two sections of the form. Please don’t close the window after you finish it. 

[7 min] [participant finishes the quiz and the survey]

\textbf{[3 min] System Introduction}

[3 min] [researcher introduces the system to the participant]

Welcome to the StopGap system. This prototype is designed to display Audiovisuals with real-time jargon word explanations in various formats. Let me upload a Audiovisual to illustrate more about the system. 
[researcher uploads the Audiovisual]
[Optional] It may take a few seconds for the Audiovisual to be uploaded. After it’s fully loaded, this play button will turn blue.
Now let’s use this play button to start. 

[researcher clicks the play button]

Now the Audiovisual starts playing, and explanations for any detected jargon words will appear on the right side in real time. In the flashcard, you can see the jargon word and its corresponding explanations in different formats. We have a total of 4 formats: text-based definition, metaphor, list, and image. The format shown here is selected by AI. While watching the Audiovisual, you can pause the Audiovisual at any time using the “Pause” button at the bottom. 

[researcher clicks the pause button]

The explanation will be paused at the same time. 
Do you have any questions or comments on the system?

\textbf{[25-28min] Experimental Task: Watch Audiovisual with StopGap assistance}

\textit{Goal: observe participants interact with the system and explore their preferred knowledge representation format}

Now, we will show you another Audiovisual. Your task is to watch and understand its content with the help of the StopGap system. Note that please do not use any external tools other than StopGap in this task. You have up to 6 minutes, including the Audiovisual length (three minutes), to pause the Audiovisual and check the explanations. There will be a short quiz and a survey after the Audiovisual ends.

[3-6 min] [researcher plays the Audiovisual; participant watches with StopGap]

Q. On a scale of 1 to 5, with 1 being very easy and 5 being very difficult, how would you rate the difficulty of understanding this Audiovisual? 

Q. How did StopGap impact your understanding?

Please go back to the quiz and finish it, again, you’ll have up to 7 minutes

[7 min] [participant finishes the quiz and the survey]

We have different formats for explaining the jargon words, such as text-based definitions, lists, metaphors, and images. For each jargon word, you only saw one format in the system. In this task, we’re going to walk you through the jargon words you just saw and all of the corresponding formats. Please let us know which format you find most useful in enhancing your understanding in real-time settings and elaborate on your choice. In other words, we want to know whether you want to replace the current explanation format in the system with any others. 

[10 min] [researcher walks through all representations for each jargon word]
\begin{enumerate}
    \item For each jargon word, explain their choice.
    \item What about the other formats? Any comments and thoughts?
\end{enumerate}	
\quad [Overall comment] When would you use a specific format? 

\textbf{[15 min]: Debriefing Interview}

\textit{Goal: To compare their experiences across the two tasks and talk about the StopGap and the future of real-time knowledge support systems}

[10 min] [researcher asks about experiences watching the Audiovisuals with and without the system]
\begin{enumerate}
    \item Could you compare the benefits of with and without the support of StopGap?
    \item Could you compare the difficulties you encountered with and without StopGap support?
    \item Which do you prefer and why?
\end{enumerate}
[10 min] [researcher asks open-ended questions on provided real-time knowledge support]
\begin{enumerate}
    \item (if the participant paused the Audiovisual) why did you pause? [Ask about the different points – give context based on the jargon word there were on] 
    \item From 1-5, with 1 being not useful at all, and 5 being extremely useful, how useful was the provided knowledge support? Why?
    \item What do you like and dislike about the system?
    \item Do you have any suggestions for improvements for such real-time knowledge support tools?
\end{enumerate}


\section{StopGap In-Study Questionnaire}
\label{appendix:questionnaire}
We investigated participants' understanding of the video content and cognitive load while watching the video with and without the StopGap system using this questionnaire.

\subsection{Video 1 Quiz (Without Using StopGap System)}

\begin{enumerate}
    \item What is the significance of proper gene segregation during cell division?
    \begin{enumerate}
        \item It helps increase energy production in cells.
        \item It ensures each daughter cell receives the correct number of genes.
        \item It enhances the immune response.
        \item It improves protein synthesis in cells.
    \end{enumerate}
    
    \item What role do microtubules play in cell division?
    \begin{enumerate}
        \item They store genetic information.
        \item They build a spindle-like structure to segregate duplicated genes.
        \item They enhance nutrient absorption.
        \item They increase cell metabolism.
    \end{enumerate}

    \item Why is fluorescence microscopy used in studying cell division?
    \begin{enumerate}
        \item It measures the pH levels within cells.
        \item It helps visualize dynamic microtubules live in action.
        \item It increases the resolution of electron microscopy.
        \item It enhances protein synthesis in cells.
    \end{enumerate}

    \item What happens if the spindle structure does not function correctly during cell division?
    \begin{enumerate}
        \item The cell increases its energy production.
        \item The cell pauses division until the orientation is fixed.
        \item The cell absorbs more nutrients.
        \item The cell immediately dies without any attempt to fix the issue.
    \end{enumerate}
\end{enumerate}

\subsubsection{Familiarity with Jargon Words}
\begin{enumerate}
    \item How familiar are you with the jargon words before watching the video?
    \begin{itemize}
        \item Not familiar at all
        \item Slightly familiar
        \item Neutral
        \item Very familiar
        \item Extremely familiar
    \end{itemize}

    \item How familiar are you with the jargon words after using your preferred method?
    \begin{itemize}
        \item Not familiar at all
        \item Slightly familiar
        \item Neutral
        \item Very familiar
        \item Extremely familiar
    \end{itemize}
\end{enumerate}

\subsubsection{Task Load Index for Video 1}
\begin{itemize}
    \item On a scale of 1 to 21, how mentally demanding was it for you to watch the video and understand the jargon words? (1 being very low and 21 being very high)
    \item On a scale of 1 to 21, how stressed or annoyed did you feel while watching the video and trying to understand the jargon words? (1 being very low and 21 being very high)
    \item On a scale of 1 to 21, how hurried or rushed did you feel while watching the video and trying to understand the jargon words? (1 being very low and 21 being very high)
    \item On a scale of 1 to 21, how successful were you in understanding the jargon words in the video? (1 being very low and 21 being very high)
    \item On a scale of 1 to 21, how much effort did you have to exert to understand the jargon words in the video? (1 being very low and 21 being very high)
\end{itemize}

\subsection{Video 2 Quiz (Using StopGap System)}

\begin{enumerate}
    \item What is the main drawback of traditional chemotherapy?
    \begin{enumerate}
        \item It enhances protein synthesis in all cells.
        \item It increases the immune response in cancer patients.
        \item It kills more harmless cells as collateral damage relative to cancer cells.
        \item It provides nutrients to cancer cells.
    \end{enumerate}

    \item What role do tubulin genes play in the cell?
    \begin{enumerate}
        \item They enhance energy production.
        \item They provide structure and support, similar to a skeleton in the human body.
        \item They store genetic information.
        \item They regulate nutrient absorption.
    \end{enumerate}

    \item Why are microtubule systems important in cancer research?
    \begin{enumerate}
        \item They enhance the immune response in cancer patients.
        \item They are composed of multi-colored subunits that represent different isotypes of tubulin, helping in cell division.
        \item They increase protein synthesis in cancer cells.
        \item They measure pH levels within cancer cells.
    \end{enumerate}

    \item What is the significance of understanding the language of tubulin genes?
    \begin{enumerate}
        \item It helps in improving the immune response.
        \item It provides insights into nutrient absorption.
        \item It aids in developing novel therapeutic targets against cancer.
        \item It increases the metabolic rate of cancer cells.
    \end{enumerate}
\end{enumerate}

\subsubsection{Familiarity with Jargon Words}
\begin{enumerate}
    \item How familiar are you with the jargon words before watching the video?
    \begin{itemize}
        \item Not familiar at all
        \item Slightly familiar
        \item Neutral
        \item Very familiar
        \item Extremely familiar
    \end{itemize}

    \item How familiar are you with the jargon words after using the system?
    \begin{itemize}
        \item Not familiar at all
        \item Slightly familiar
        \item Neutral
        \item Very familiar
        \item Extremely familiar
    \end{itemize}
\end{enumerate}

\subsubsection{Task Load Index for Video 2}
\begin{itemize}
    \item On a scale of 1 to 21, how mentally demanding was it for you to watch the video and understand the jargon words with StopGap? (1 being very low and 21 being very high)
    \item On a scale of 1 to 21, how stressed or annoyed did you feel while watching the video and trying to understand the jargon words with StopGap? (1 being very low and 21 being very high)
    \item On a scale of 1 to 21, how hurried or rushed did you feel while watching the video and trying to understand the jargon words with StopGap? (1 being very low and 21 being very high)
    \item On a scale of 1 to 21, how successful were you in understanding the jargon words in the video? (1 being very low and 21 being very high)
    \item On a scale of 1 to 21, how much effort did you have to exert to understand the jargon words in the video? (1 being very low and 21 being very high)
\end{itemize}

\section{Comparison of Self-reported TLX between Audio and Audiovisual Groups}
\label{appendix:compare}
\begin{figure}[h]
    \centering
    \begin{subfigure}[b]{0.49\textwidth}
        \centering
        \includegraphics[width=\linewidth]{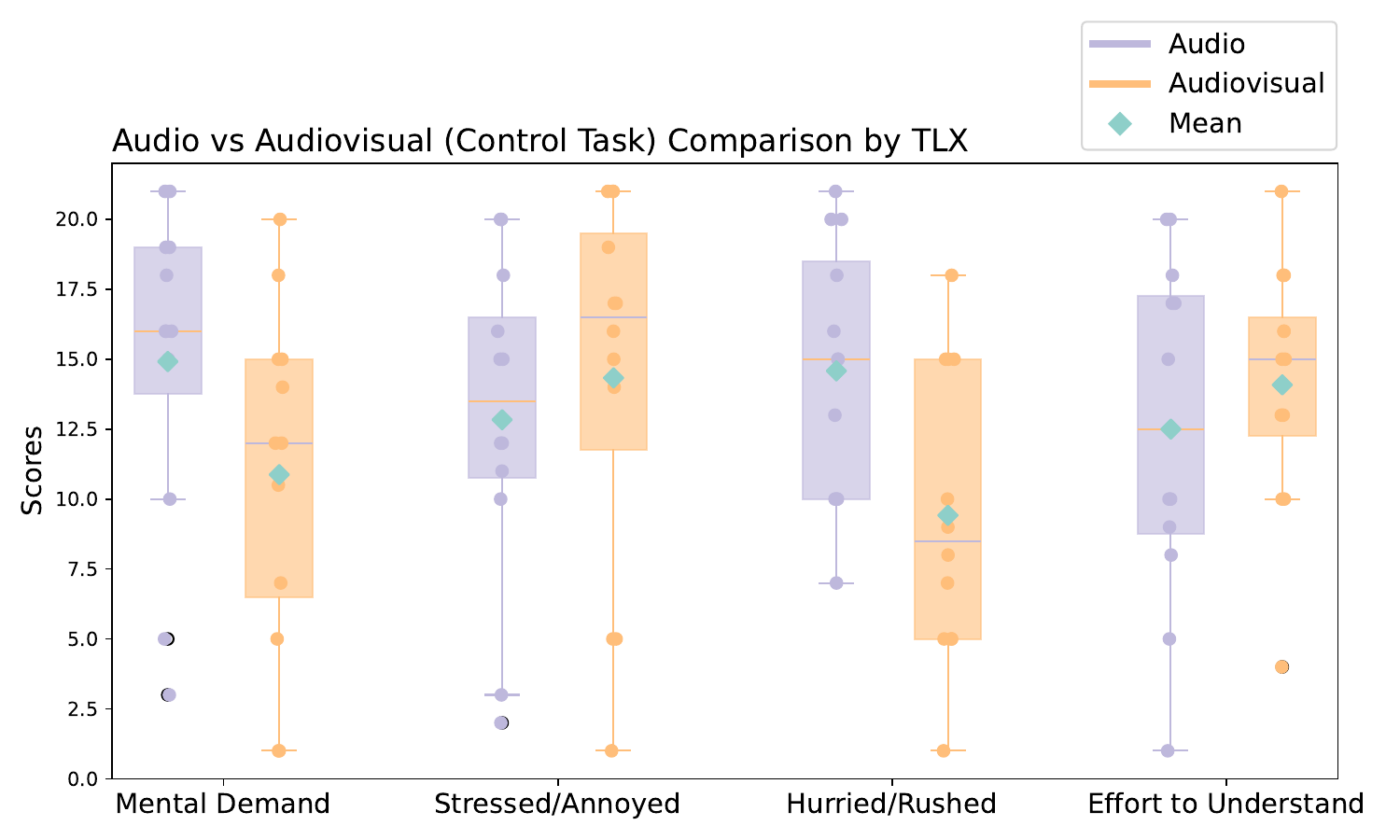}
        \caption{Comparison of participants' task load index between audio and audiovisual groups of the control task}
        \label{fig:control_tlx}
    \end{subfigure}
    \hfill
    \begin{subfigure}[b]{0.49\textwidth}
        \centering
        \includegraphics[width=\linewidth]{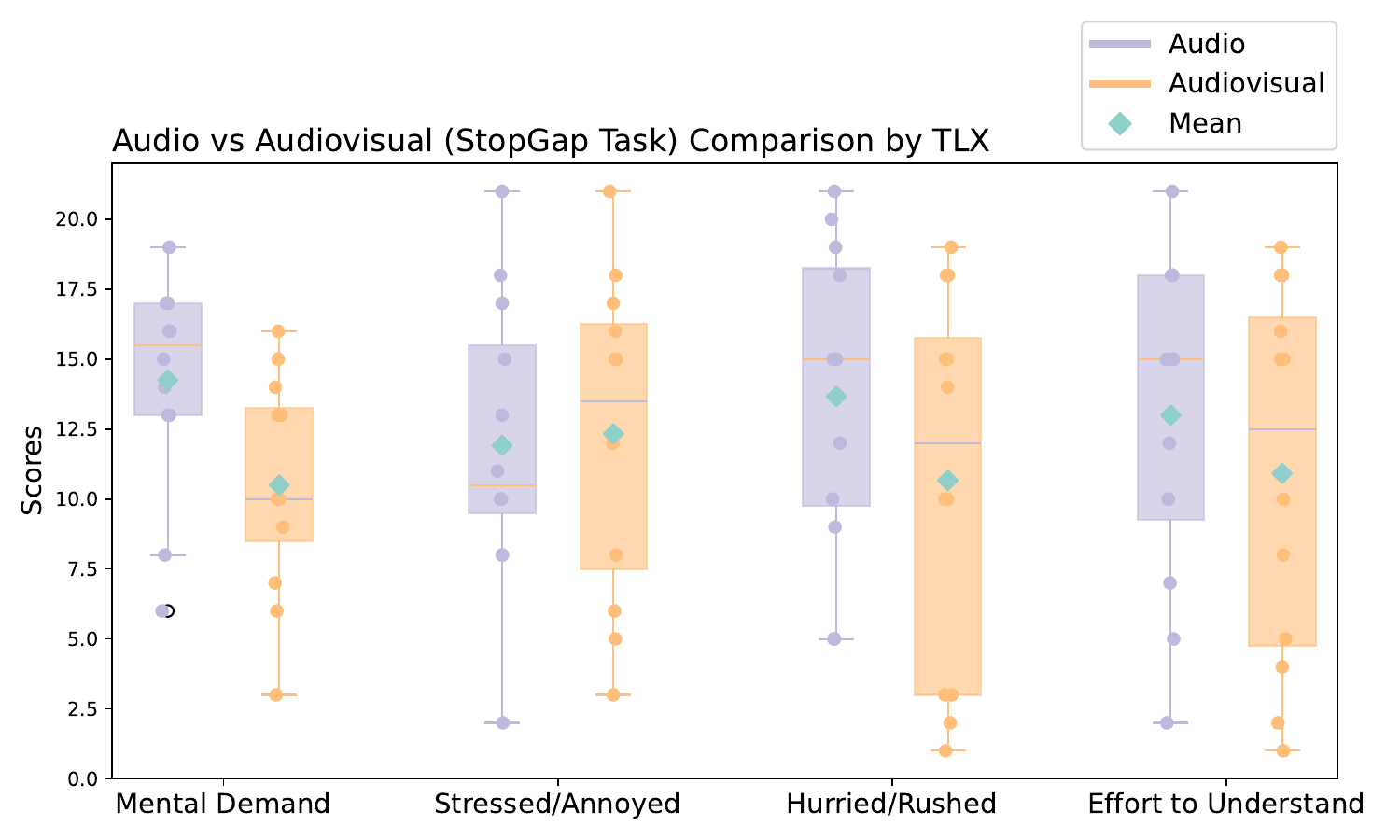}
        \caption{Comparison of participants' task load index between audio and audiovisual groups of the \StopGap task}
        \label{fig:stopgap_tlx}
    \end{subfigure}
    
    \caption{Comparison of participants' cognitive load between audio and audiovisual groups for both control and \StopGap tasks.}
    \label{fig:tlx_comparison}
    \Description{Figure 7 a): Comparison of participants' task load index in the control task. Audio group has higher mean in mentally demand and hurried or rushed while audiovisual group has higher mean in the other two. Distribution in audio group is more concentrated in all metrics except effort to understand.
Figure 7 b): Comparison of participants' task load index in the StopGap task. Audio group has higher mean in all metrics except stressed or annoyed, where in that metric two groups are very similar. Distribution in audio group is more concentrated in all metrics except mentally demand.
}
\end{figure}

\section{Jargon Word Explanations}
\label{appendix:explanations}
All representation formats are shown to participants in the reflective interview and the highlighted ones are displayed in the system in \StopGap task
\begin{figure}[h]
    \includegraphics[width=\linewidth]{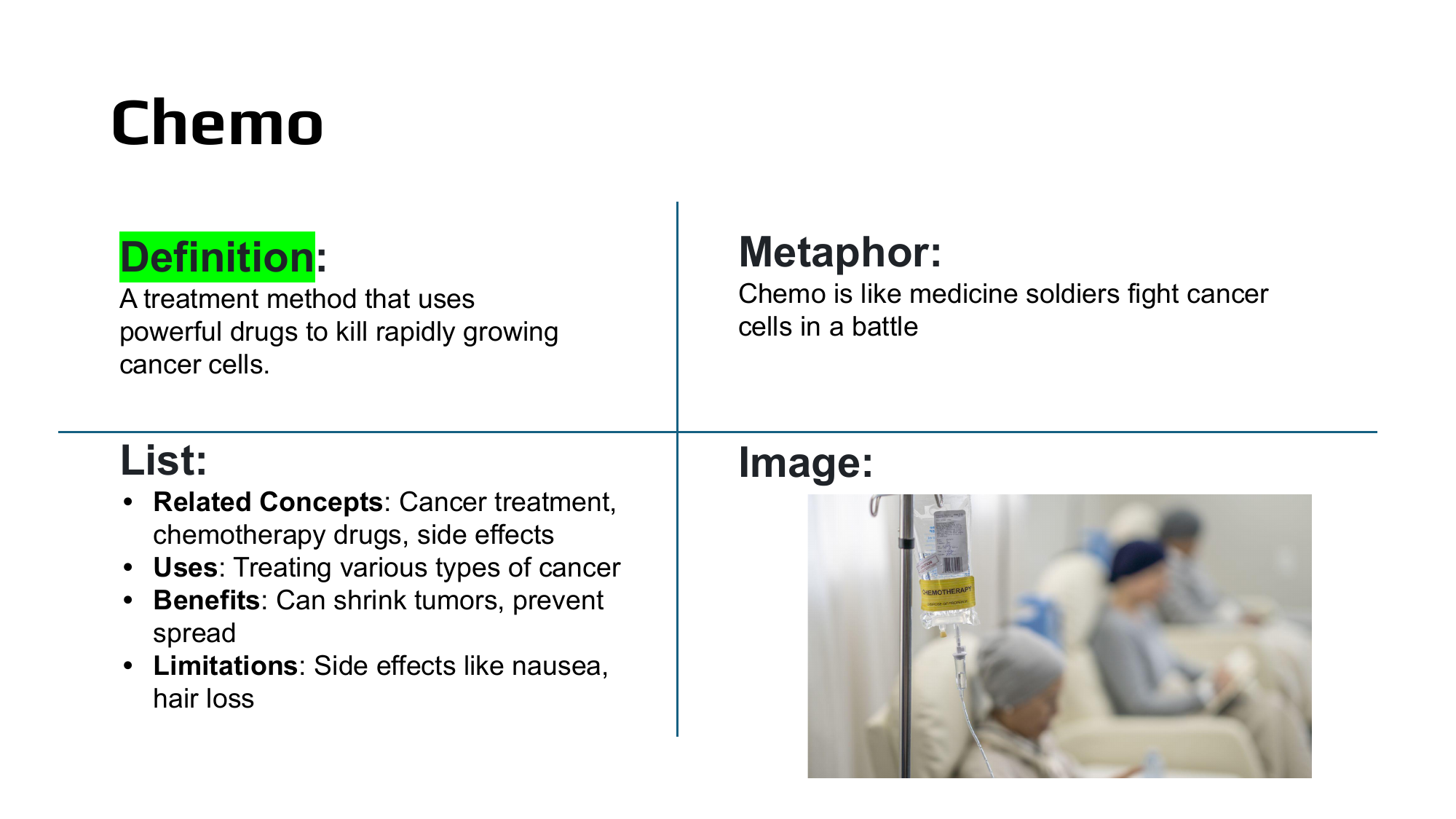}
\end{figure}
\begin{figure}[h]
    \includegraphics[width=\linewidth]{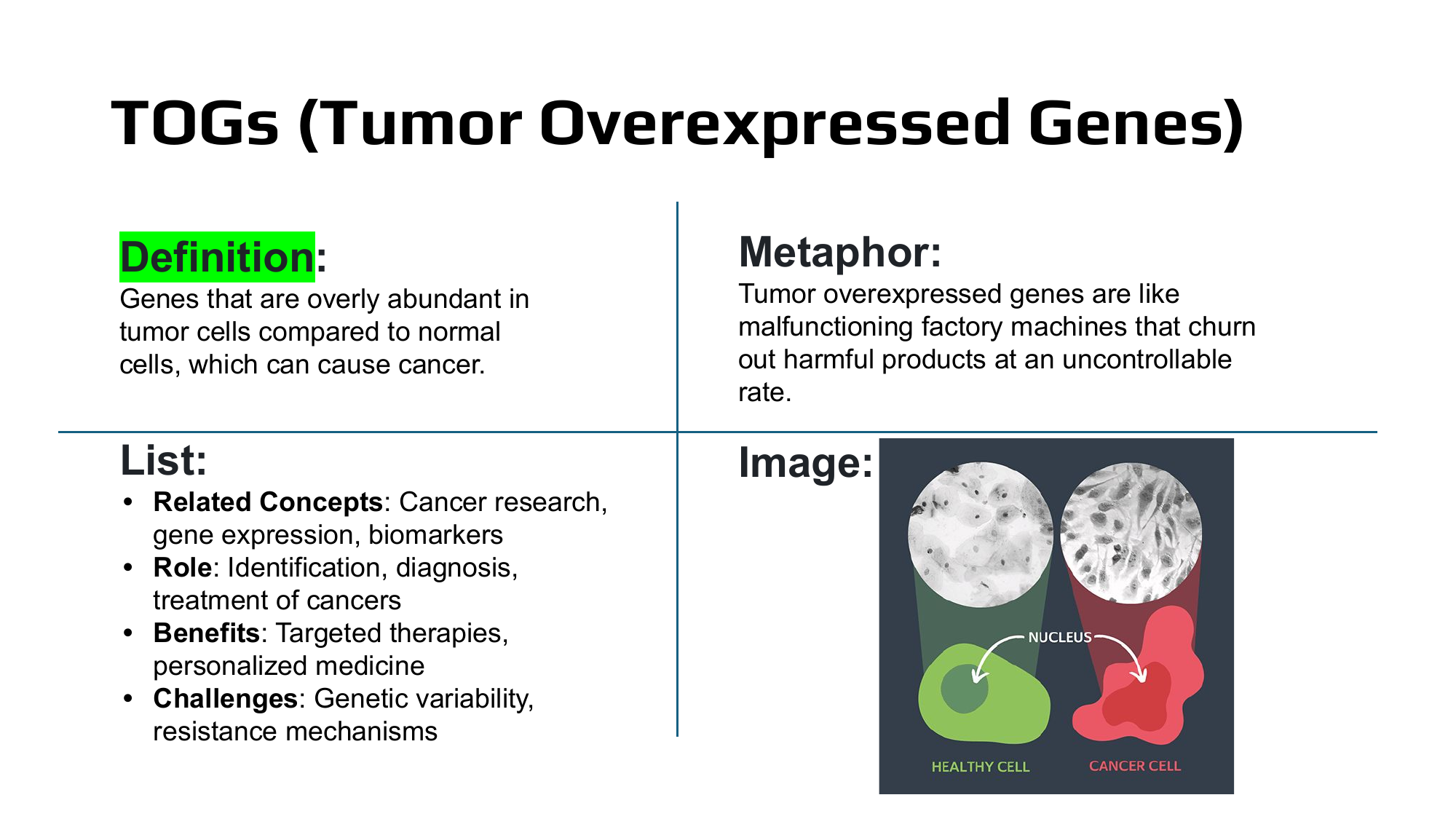}
\end{figure}
\begin{figure}[h]
    \includegraphics[width=\linewidth]{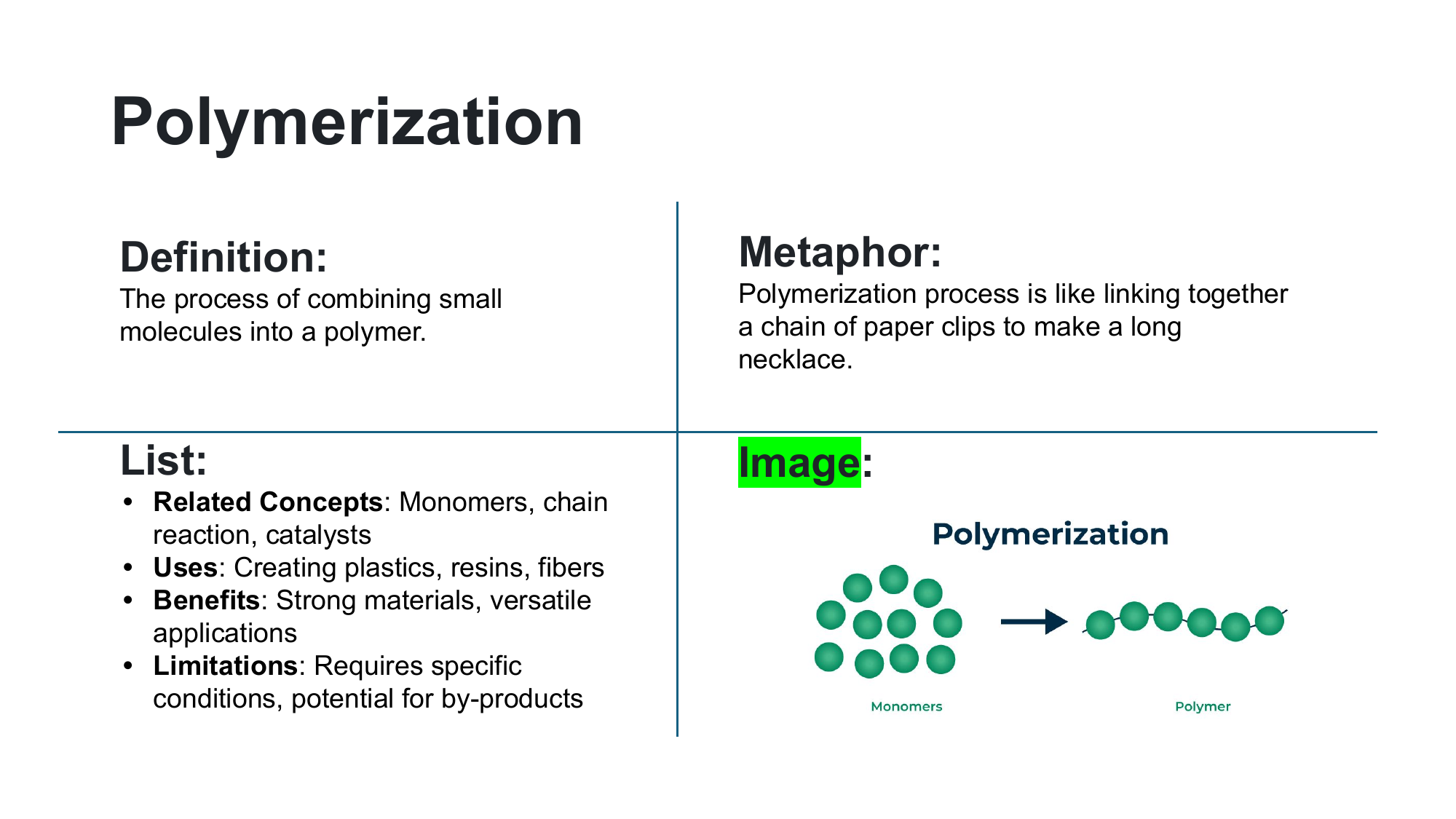}
\end{figure}
\begin{figure}[h]
    \includegraphics[width=\linewidth]{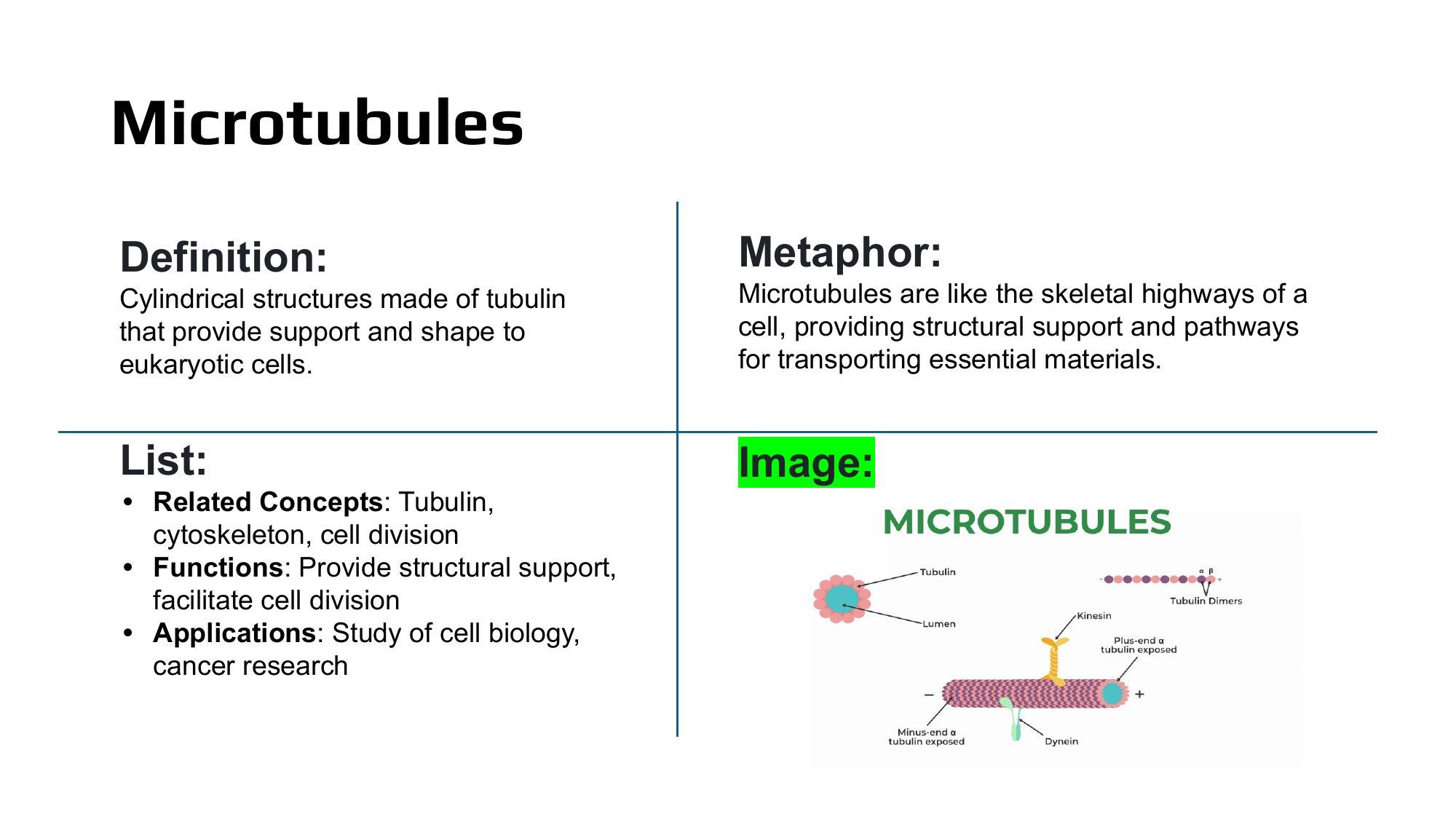}
\end{figure}
\begin{figure}[h]
    \includegraphics[width=\linewidth]{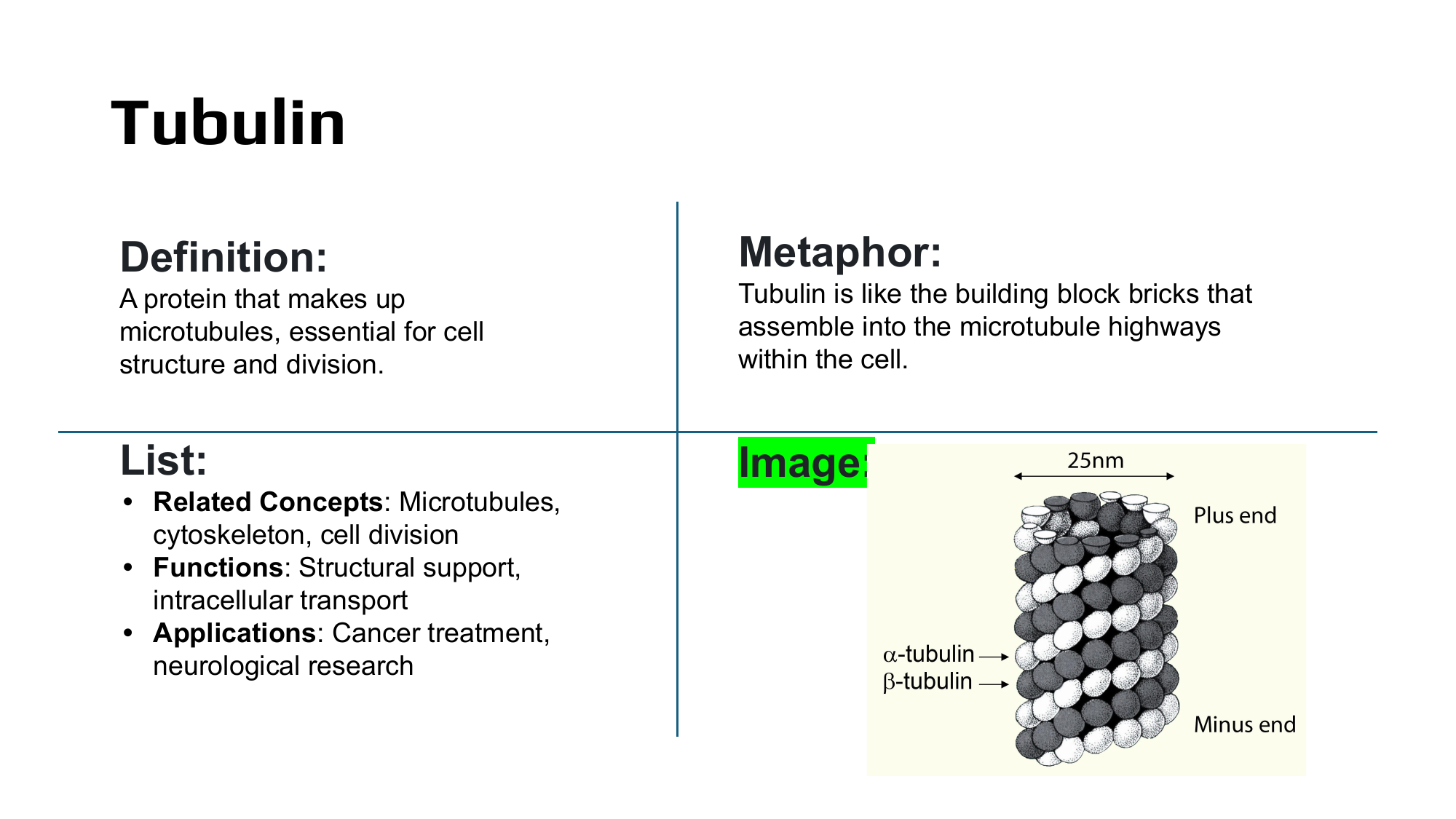}
\end{figure}
\begin{figure}[h]
    \includegraphics[width=\linewidth]{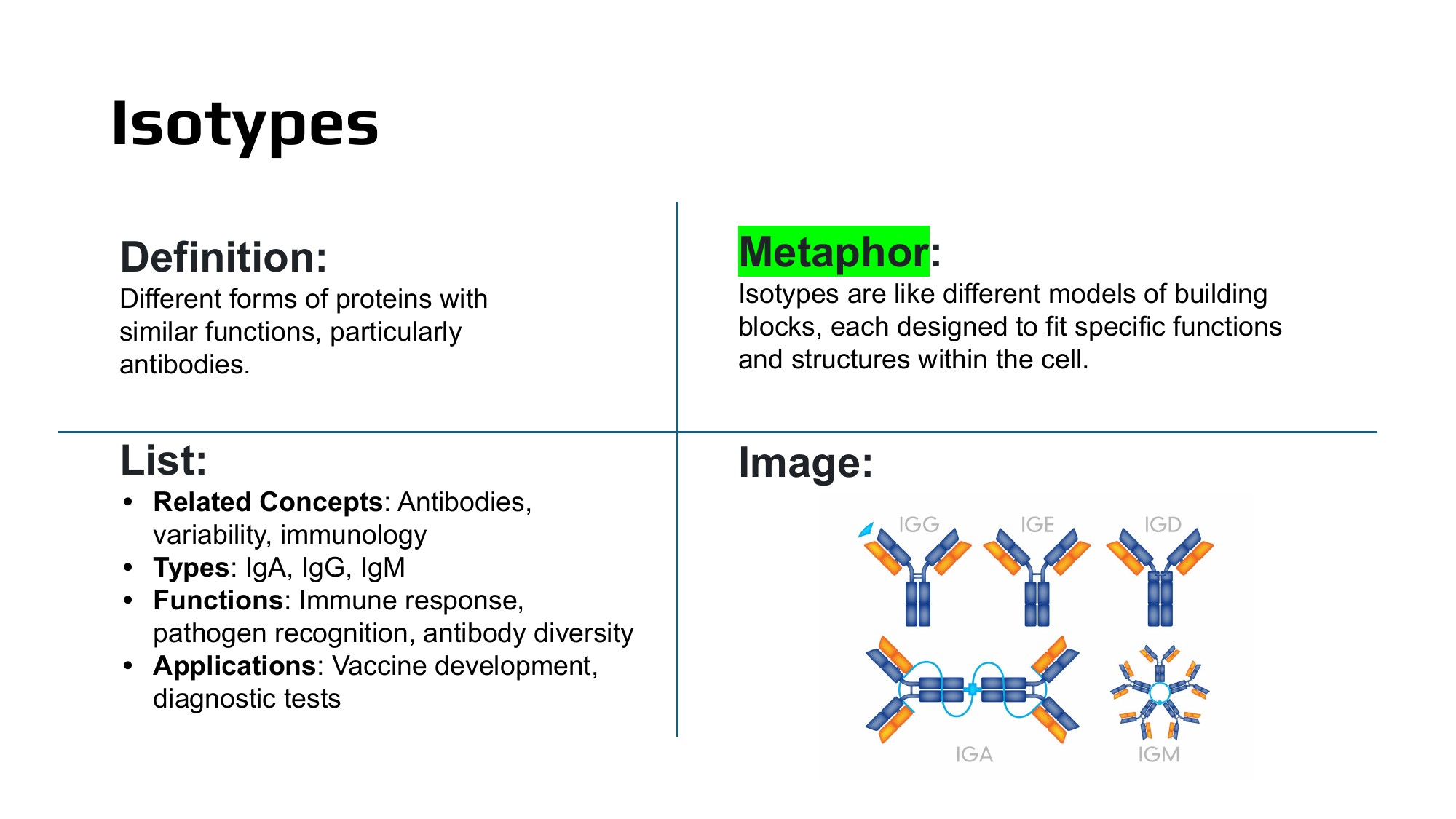}
\end{figure}
\begin{figure}[h]
    \includegraphics[width=\linewidth]{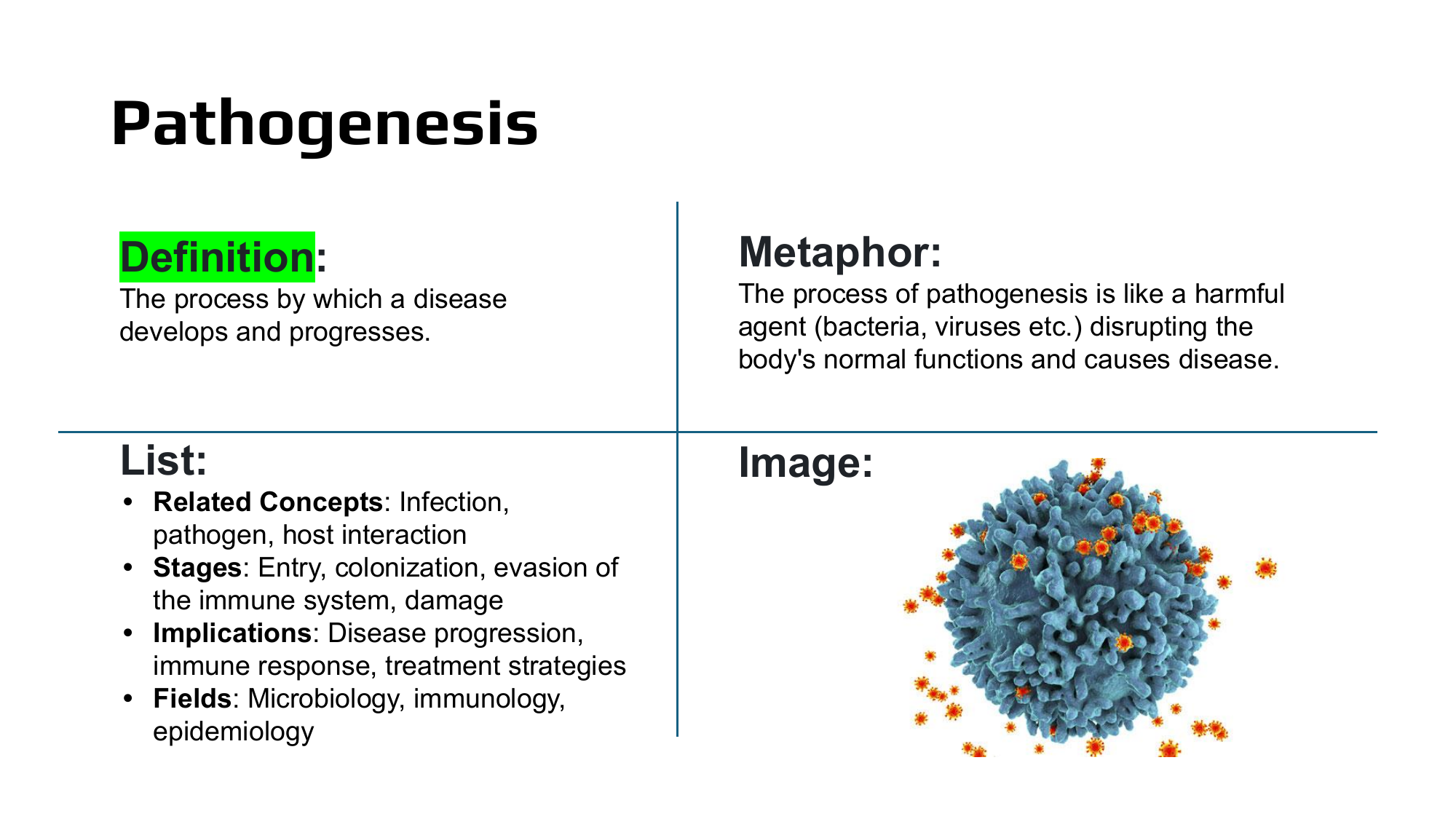}
\end{figure}
\begin{figure}[h]
    \includegraphics[width=\linewidth]{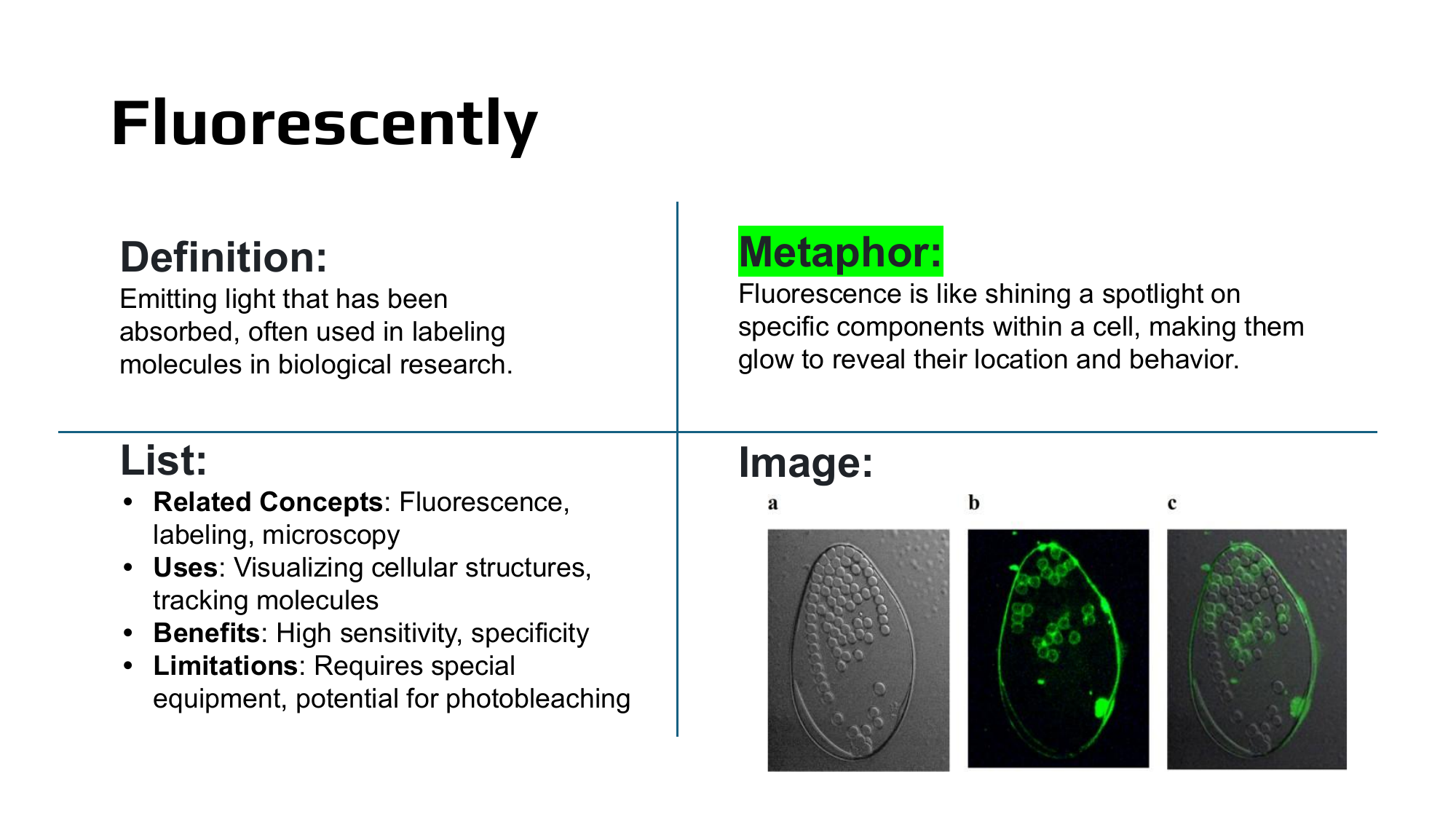}
\end{figure}
\begin{figure}[h]
    \includegraphics[width=\linewidth]{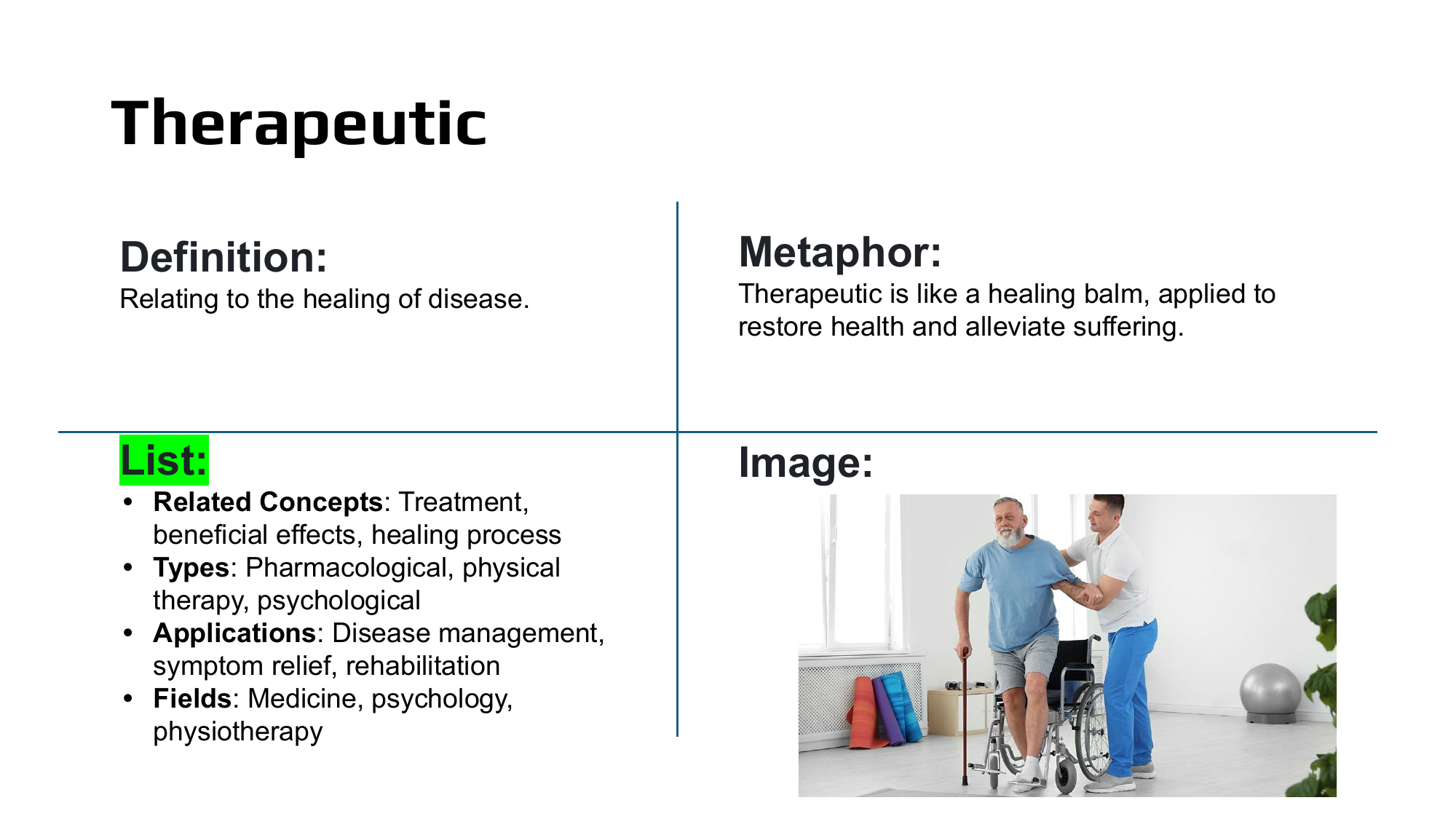}
\end{figure}

\clearpage
\section{Participant Details}
\label{appendix:participants}
\begin{table}[h]
    \centering
    \caption{Participant Demographics. Note that * indicates a counterbalancing setting where participants conducted \StopGap task before the control task.}
    \begin{tabular}{lllllll}
    \hline
\textbf{ID} & \textbf{Group} & \textbf{Age} & \textbf{Gender} & \makecell{\textbf{Professional} \\ \textbf{Background}} & \textbf{Education Background} & \makecell{\textbf{Familiarity w/} \\ \textbf{Biology}} \\ \hline
        P1 & Audio & 18-24 & Woman & 1-3 years & BS in Media Culture and Communications & Slightly familiar \\ \hline
        P2 & Audio* & 18-24 & Man & 1-3 years & BS In Quantitative Science & Not familiar at all \\ \hline
        P3 & Audiovisual & 25-34 & Man & 1-3 years & BS in System Engineering & Slightly familiar \\ \hline
        P4 & Audiovisual* & 25-34 & Woman & 1-3 years & BS in Human \& Organizational Development & Slightly familiar \\ \hline
        P5 & Audio & 18-24 & Woman & 1-3 years & BA in Political Science \& Economics  & Slightly familiar \\ \hline
        P6 & Audio* & 25-34 & Woman & 4-5 years & BS in Industrial Engineering & Slightly familiar \\ \hline
        P7 & Audiovisual & 18-24 & Man & 1-3 years & BS in Computer Science and Engineering & Not familiar at all \\ \hline
        P8 & Audiovisual* & 35-44 & Man & 10+ years & MS in Accounting and MBA & Not familiar at all \\ \hline
        P9 & Audio & 18-24 & Woman & <1 year & BS in Computer Science & Not familiar at all \\ \hline
        P10 & Audio* & 25-34 & Man & 4-5 years & BA in Government & Slightly familiar \\ \hline
        P11 & Audiovisual & 18-24 & Woman & 1-3 years & BS in Managerial Economics & Slightly familiar \\ \hline
        P12 & Audiovisual* & 45-54 & Woman & 10+ years & BA in in English Literature & Not familiar at all \\ \hline
        P13 & Audio & 35-44 & Man & 10+ years & MS in Electrical and Optical Engineering & Slightly familiar \\ \hline
        P14 & Audio* & 18-24 & Woman & 1-3 years & BA in Statistics and Data Science & Neutral \\ \hline
        P15 & Audiovisual & 25-34 & Woman & 1-3 years & BS in Business Administration & Not familiar at all \\ \hline
        P16 & Audiovisual* & 35-44 & Man & 10+ years & BE in Civil Engineering & Neutral \\ \hline
        P17 & Audio & 25-34 & Man & 1-3 years & BS in Business Administration & Neutral \\ \hline
        P18 & Audio* & 25-34 & Woman & 4-5 years & BA in Economics & Neutral \\ \hline
        P19 & Audiovisual & 18-24 & Man & <=1 years & BS Industrial Engineering \& Operational Research & Slightly familiar \\ \hline
        P20 & Audiovisual* & 18-24 & Woman & 1-3 years & BA in Economics and Accounting & Slightly familiar \\ \hline
        P21 & Audio & 18-24 & Man & 1-3 years & BS in Computer Science & Slightly familiar \\ \hline
        P22 & Audio* & 25-34 & Woman & 1-3 years & BS in Managerial Economics & Slightly familiar \\ \hline
        P23 & Audiovisual & 45-54 & Man & 10+ years & BS in Mathematics and Computer Science & Not familiar at all \\ \hline
        P24 & Audiovisual* & 18-24 & Woman & 4-5 years & BS in Management Information Systems & Slightly familiar \\ \hline
    \end{tabular}
\end{table}



\end{document}